%% file: main.tex
\documentclass[conference,letterpaper]{IEEEtran}
\usepackage[english]{babel}
\usepackage[utf8x]{inputenc}
\usepackage[T1]{fontenc}
\usepackage{tikz}
\usepackage{graphicx}
\usetikzlibrary{positioning}
\usepackage{caption}
\usepackage{subcaption}
\usepackage{todonotes}
\usepackage{hyperref}
\usepackage{listings}
\usepackage{xspace}
\usepackage{ifthen}
\usepackage{comment}
\usepackage{amsmath}
\usepackage{cleveref}
\usepackage{ulem}

\usepackage{filecontents,lipsum}
\usepackage[noadjust]{cite}

\renewcommand{\emph}{\textit}
\renewcommand{\paragraph}[1]{\noindent\textbf{#1}}

\usepackage[T1]{fontenc}

\graphicspath{{./}}

\usepackage{acro}
\acsetup{first-long-format=\itshape}
\include{acronyms}

\newcommand{\name}{\textsc{Scadman}\xspace}

\newcommand{\masterplc}{\textsc{Scadman-Monitor}\xspace}
\newcommand{\ltoiec}{\textsc{L5X2IEC}\xspace}

\newcommand{\plc}[1]{$\textrm{PLC#1}$\xspace}

\newcommand{\says}[3]{\todo[size=\small,color=#2,inline]{#1 says: #3}}

\newcommand{\michael}[1]{\says{Michael}{yellow}{#1}}

\newcommand{\green}[1]{#1}

\newcommand{\red}[1]{#1}

\begin{document}

\author{
\IEEEauthorblockN{Sridhar Adepu$^{*1}$, Ferdinand Brasser$^{\dagger2}$, Luis Garcia$^{\ddagger3}$, Michael Rodler$^{\oplus4}$, \\Lucas Davi$^{\oplus5}$, Ahmad-Reza Sadeghi$^{\dagger6}$, Saman Zonouz$^{\circ7}$}
\IEEEauthorblockA{$^*$Singapore University of Technology and Design
Singapore;  $^\dagger$Technische Universit{\"a}t Darmstadt;\\ $^\ddagger$University of California, Los Angeles; $^\oplus$University of Duisburg-Essen; $^\circ$Rutgers University\\
adepu@sutd.edu.sg$^1$; \{ferdinand.brasser$^2$,ahmad.sadeghi$^6$\}.trust.tu-darmstadt.de; garcialuis@ucla.edu$^3$; \\\{michael.rodler$^4$, lucas.davi$^5$\}@uni-due.de szonouz@rutgers.edu$^7$}

}
\IEEEoverridecommandlockouts

	\title{Control Behavior Integrity for Distributed Cyber-Physical Systems}
    \maketitle
	
    \input{0-abstract}
    \input{1-introduction}
    \input{2-background}

    \input{3-systemmodel}
    \input{4-framework}

    \input{5-llvm-instrumentation}

    \input{6-security}
    \input{7-evaluations}
    \input{8-related-work}

    \input{9-discussions}
    \input{10-conclusions}

	\bibliographystyle{IEEEtranS}
	\bibliography{reference}

\end{document}

%% file: acronyms.tex
\DeclareAcronym{rop}{
  short = ROP,
  long = return-oriented programming
}
\DeclareAcronym{dop}{
  short = DOP,
  long = data-oriented programming
}
\DeclareAcronym{cfi}{
	short = CFI,
    long = control-flow integrity
}
\DeclareAcronym{ics}{
	short = ICS,
    long = industrial control system,
    long-plural = industrial control systems
}
\DeclareAcronym{plc}{
	short = PLC,
    long = programmable logic controller,
    short-plural = s,
    long-plural = s
}
\DeclareAcronym{cps}{
	short = CPS,
    long = cyber physical system,
    long-plural = s
}
\DeclareAcronym{ssa}{
	short = SSA,
    long = single static assignment
}

\DeclareAcronym{cbi}{
	short = cbi,
    long = control behavior integrity
}

\DeclareAcronym{swat}{
	short = SWaT,
    long = secure water treatment plant
}

%% file: 0-abstract.tex
\begin{abstract}

Cyber-physical control systems, such as industrial control systems (ICS), are increasingly targeted by cyberattacks. Such attacks can potentially cause tremendous damage, affect critical infrastructure or even jeopardize human life when the system does not behave as intended.
Cyberattacks, however, are not new and decades of security research have developed plenty of solutions to thwart them. Unfortunately, many of these solutions cannot be easily applied to safety-critical cyber-physical systems. Further, the attack surface of ICS is quite different from what can be commonly assumed in classical IT systems.

We present \name, a system with the goal to preserve the \emph{Control Behavior Integrity} (CBI) of distributed cyber-physical systems. By observing the system-wide behavior, the correctness of individual controllers in the system can be verified. This allows \name to detect a wide range of attacks against controllers, like programmable logic controller (PLCs), including malware attacks, code-reuse and data-only attacks.
We implemented and evaluated \name based on a real-world water treatment testbed for research and training on ICS security. Our results show that we can detect a wide range of attacks--including attacks that have previously been undetectable by typical state estimation techniques--while causing no false-positive warning for nominal threshold values.

\end{abstract}

%% file: 1-introduction.tex
\section{Introduction}
\label{sec:introduction}

\Ac{ics} are used in a multitude of control systems across several applications of industrial sectors and critical infrastructures, including electric power transmission and distribution, oil and natural gas production, refinery operations, water treatment systems, wastewater collection systems, as well as pipeline transport systems~\cite{stouffer2011guide}. 
ICS typically consist of interconnected embedded systems, called programmable logic controllers (PLCs).
In a distributed ICS, multiple PLCs jointly control a physical process or the physical environment.
Using a series of sensors and actuators, PLCs can monitor the physical system's state and control the system behavior.
This makes the correct functioning of PLCs crucial for the correct and safe operation of these systems.

This critical role of the PLCs makes them a valuable target for adversaries aiming to interfere with any of these systems~\cite{adepuMishraMathur}.
Past incidences show that such attacks are applied in practice, often remaining undetected over a long period of time.
Examples include the infamous Stuxnet worm~\cite{stuxnet-2010} against Iranian nuclear uranium enrichment facilities as well as the BlackEnergy crimeware~\cite{blackenergy} against the Ukranian train railway and electricity power industries.
These attacks demonstrate impressively that targeted attacks on critical infrastructure can evade traditional cybersecurity detection and cause catastrophic failures with substantive impact. 
The discover\green{ies} of Duqu~\cite{duqu-2011} and Havex~\cite{rrushiquantitative} show that such attacks are not isolated cases as they infected ICS in more than eight countries. 
Nation-state ICS malware \red{has} typically either targeted the control programs of PLCs or the central control infrastructure (e.g., operator workstations).
However, academic research has demonstrated even more sophisticated attacks against ICS and PLCs that can circumvent existing defense mechanisms by manipulating the PLC's firmware and incorporating physics-aware models into the attack code~\cite{garcia2017hey}.

A comprehensive defense against ICS attacks needs to protect against various attack vectors.
(1)~The software determining a PLC's behavior could be replaced by a malicious program~\cite{stuxnet-2010,KLM+15,BS15}. 
Updating the PLC control program over the network is an intended functionality of PLCs to allow central management.
However, the control program can also be manipulated if an attacker gains physical access to a PLC. 
(2)~The PLC firmware (which includes the OS) could be manipulated/replaced either via the network or through physical access~\cite{garcia2017hey,BBL+13}.
(3)~The attack can exploit a memory corruption vulnerability (e.g., buffer overflow~\cite{S14}) in the PLC's control programs and/or firmware for code-injection or to launch run-time attacks such as return-oriented programming (ROP)~\cite{rop}, to manipulate a PLC's behavior.
(4)~Memory corruption vulnerabilities can be exploited to launch data-only attacks~\cite{dop} against a PLC to manipulate its behavior.
For instance, the initiation of a trigger-response may be inhibited by manipulating the associated control parameters~\cite{robotlaws}, e.g., a threshold value that determines whether the action must be started. 

For all above enumerated attack vectors, a common goal of the adversary is to modify the physical behavior of the system.
As long as the attacked device behaves correctly the overall system will continue to operate correctly.
Therefore, the ultimate goal of the attacker will always be to change a device's behavior.

Previous works in defending against ICS attacks focus only on a subset of the above listed attack vectors. These approaches can be generally categorized into two categories: defenses that focus on verifying the integrity of the software running on a PLC and defenses that verify the behavior of the overall ICS based on models that abstract control decisions of the PLC software. In the former case, PLC-based verification solutions typically cannot account for attacks which replace and/or modify either the application layer programs or the underlying firmware. For instance, ECFI~\cite{ecfi} provides protection against run-time attacks targeting PLC control programs, but does not protect against data-only attacks nor maliciously modified/replaced control programs or firmware. Orpheus~\cite{cheng2017orpheus} monitors the behavior of a device's control program based on the invoked system calls. Attacks are detected based on a finite-state machine (FSM) representing the control program's benign system call behavior. Orpheus' behavior monitor is placed inside the device's OS, hence, a compromised OS can disable and circumvent its protection mechanism. 

Similarly, solutions that enforce compliance via state estimation~\cite{adepu2016using}  or cyber-physical access control~\cite{etigowni2016cpac} from within the PLC could be circumvented as well. Zeus~\cite{han2017watch} uses side-channel analysis to verify the \green{software} control flow of programs running on a PLC, but cannot defend against firmware modifications nor sensor data attacks. 
By extension, offline, static analysis of control programs being loaded onto PLCs~\cite{mclaughlin2014trusted}~\cite{darvas2015plcverif} provides even less run-time guarantees. For ICS-based verification techniques, it has been shown that state estimation can be used to infer the control commands issued by distributed controllers~\cite{etigownicyber} or to detect false data injection attacks~\cite{liu2011false} based on the sensor data. Such protection mechanisms may be circumvented via physics-aware attacks~\cite{garcia2017hey,garcia2014covert}. Further, supervised machine learning has been used to characterize physical invariants of the CPS~\cite{Chen-Poskitt-Sun18a}. However, such approaches depend on the training data to include all corner cases of the system execution and are not based on the control flow of the software.




We present \name, the first \emph{control behavior integrity} (CBI) solution for distributed industrial control systems.
Unlike previous state estimation approaches \name does not abstract the behavior of the cyber-components (i.e., PLCs). Instead, \name \emph{precisely} simulates the state of all PLCs.
By monitoring the input and output behavior of the entire ICS, \name can detect inconsistencies within the actions of PLCs.
To enable a global view of the entire ICS, a consolidated control program of all PLCs in the system \red{is} generated to resolve functional dependencies between individual programs. 
The consolidated control program in conjunction with a physical state estimator is used to determine a set of acceptable states \red{at any particular point in time}.
For that, \name needs means to analyze which control-flow paths are valid given the current system state.
Based on this context-aware control-flow path analysis, \name determines benign resulting states.
Comparing the set of benign states against the reported sensor readings and actuation commands from the ICS allows \name to detect anomalies in the system behavior.
This makes \name agnostic to the \red{various} attack technique \red{listed above, that can be} used to cause a PLC to deviate from its intended behavior
and makes \name a powerful tool to protect ICS against a wide range of attack vectors.

We evaluated \name on real-world industrial control system equipment~\cite{swatDataset}, which are the quasi-standard for security research validation in the context of ICS~\cite{junejo2016data,lin2018tabor,Chen-Poskitt-Sun18a,Chen-Poskitt-Sun16a,gohAdepuTanLee,KongLiChenSunSunWang,wang2017should,WangChenSunQin,InoueYCP017,UmerMathurJunejoAdepu,PalAdepuGoh}. Simulation-based evaluation does not provide a viable option for \name. In general, simulations are based on models of an ICS similar to \name. Hence, such an evaluation would validate the accuracy of our models against the model of the simulator--leading to no meaningful results.

We make the following contributions:
\begin{itemize}
	\item We present \name, the first control behavior integrity system for distributed ICS based on a model comprising cyber \emph{and} physical components.
	\item Our solution does not require any changes to the hardware or software of the PLCs, making it independent of the PLC manufacturers. Furthermore, leaving the PLCs unmodified is important for safety certifications to remain valid in the presence of \name.
	\item We provide an automated solution that allows \name to consolidate the control programs of all PLCs in an ICS. This allows us to comprehensively simulate the control behavior of the entire system, which is important for detecting inconsistent behavior across the borders of individual PLCs. 
	\item We implemented \name using the MATIEC compiler from the OpenPLC project for automated generation of the consolidated PLC control program and LLVM for instrumentation of the consolidated PLC control program.
    \item We evaluated \name on real-world ICS network equipment. The results were very promising. Using its runtime control behavior monitoring, \name was able to detect all the attacks of different types against the platform in a time manner. 
\end{itemize}

The rest of the paper is structured as follows.
First, we provide background on the most relevant topics related to our work (industrial control systems, control-flow integrity and cyber-physical system modeling) in \Cref{sec:background}.
We define the assumptions and system model underlying our work in \Cref{sec:model}.
In \Cref{sec:design} we explain the design and main ideas of \name.
We detail on our implementation in \Cref{sec:impl}.
In \Cref{sec:security} we discuss \name's security for various attack scenarios and present our evaluation results in \Cref{sec:eval}.
Relevant related work is discussed in \Cref{sec:related-work}.
\Cref{sec:discussion} proposes future work directions, while \Cref{sec:conclusions} concludes.

%% file: 2-background.tex
\section{Background}
\label{sec:background}


In this section we first provide background on industrial control systems (ICS) in general.
Afterwards we introduce two concepts--control-flow integrity (CFI) and cyber-physical systems modeling--which have been used in the past in an attempt to secure ICS, however, none of them are sufficient to solve this challenge.

\noindent \textbf{Industrial Control Systems.}
Programmable logic controllers (PLC) are cyber-physical systems that are used to control industrial appliances.
PLCs feature input and output modules, which translate physical inputs -- in most cases current on a wire--into digital values and vice versa, to interact with the physical appliances like sensors and actuators. 



PLCs can convert sensor readings into digital values, process the readings with the built-in computing unit, and forward the outputs to actuators to manipulate the physical world.
Based on the available information about the system state, a PLC calculates the next actuations to steer the system towards a desired state.
The program running on the PLC, called \emph{control logic}, \red{defines} the control algorithm used to decide actuations.
The control logic program(s) of a PLC are Turing complete and programmable using the development environments provided by the PLC manufacturers.
The target system state towards which the PLC is working can be fixed in the control logic or could be set dynamically over the network by the ICS operator.

Control logic programs can be loaded onto PLCs and run on top of a privileged software layer like a real-time operating system (RTOS).
This privileged software layer contained in the PLC's firmware provides services to the control logic programs (e.g., networking, storage) and manages the programs' updates and execution.
The control logic programs are executed repeatedly in fixed intervals, called \emph{scan cycles}.
Furthermore, in a distributed ICS, the physical process is jointly controlled by multiple PLCs. To do so, PLCs are usually connected through a computer network, allowing them to share \red{information like} sensor readings or internal states.

The PLCs in an ICS are usually managed and monitored through central management systems, called \emph{Supervisory Control and Data Acquisition} (SCADA).
Typical components of a SCADA system are \emph{historians}, which are databases logging data from all control devices in the ICS, IT infrastructure  servers that connect the ICS to other systems such as a supply chain management system, \emph{human machine interfaces} (HMI), which allow an operator to interactively control the system, and operator workstations \green{that provide} interactive control as well as PLC reprogramming.

\red{The control logic running on the PLCs is highly application specific and is usually programmed by the plant operator itself. In particular, while the firmware and development tools for PLCs are usually closed source, the operator has full access to the control logic of the PLCs.}
\red{In order to design and setup a control system, the operator usually needs knowledge, i.e., a model, of the physical processes in the system.}



\noindent \textbf{Control-Flow Integrity.}
Control-flow integrity (CFI) is a defense mechanism against run-time attacks.
Modern run-time attacks do not inject or modify the code of a system. Instead, they reuse the existing code by hijacking the control flow of a program in order to cause unintended, malicious program behavior~\cite{rop}.
These attacks have been demonstrated on various platforms and devices, including embedded architectures like ARM~\cite{rop-arm}, SPARC~\cite{rop-sparc} and Atmel AVR~\cite{rop-avr}.

CFI enforces that a program's control flow does not deviate from the developer-intended flow.
The integrity of the program flow is ensured by validating for each control-flow decision if the executed path lies within the program's control-flow graph~\cite{ABE+05,ABE+09}. 

In the context of ICS, the guarantees provided by CFI are not sufficient.
In particular, data-only attacks like data-oriented programming (DOP)~\cite{dop} pose a severe threat to PLCs.
Simple modifications like changing a threshold value can have catastrophic consequences, e.g., in the attack against a steel-mill, the blast furnace cloud not be turned off due to compromised controllers, resulting in massive damage.\footnote{\url{https://www.wired.com/2015/01/german-steel-mill-hack-destruction/}}


\noindent \textbf{Cyber-Physical Systems Modeling.} ICS comprise a class of cyber-physical systems that can be modeled as \emph{hybrid systems}, or systems whose continuous evolution (physical equations) evolve based on the discrete-state transitions (controller actuations) of the system~\cite{antsaklis1993hybrid}. 

\begin{figure}[th]
	\centering
	\includegraphics[width=\linewidth]{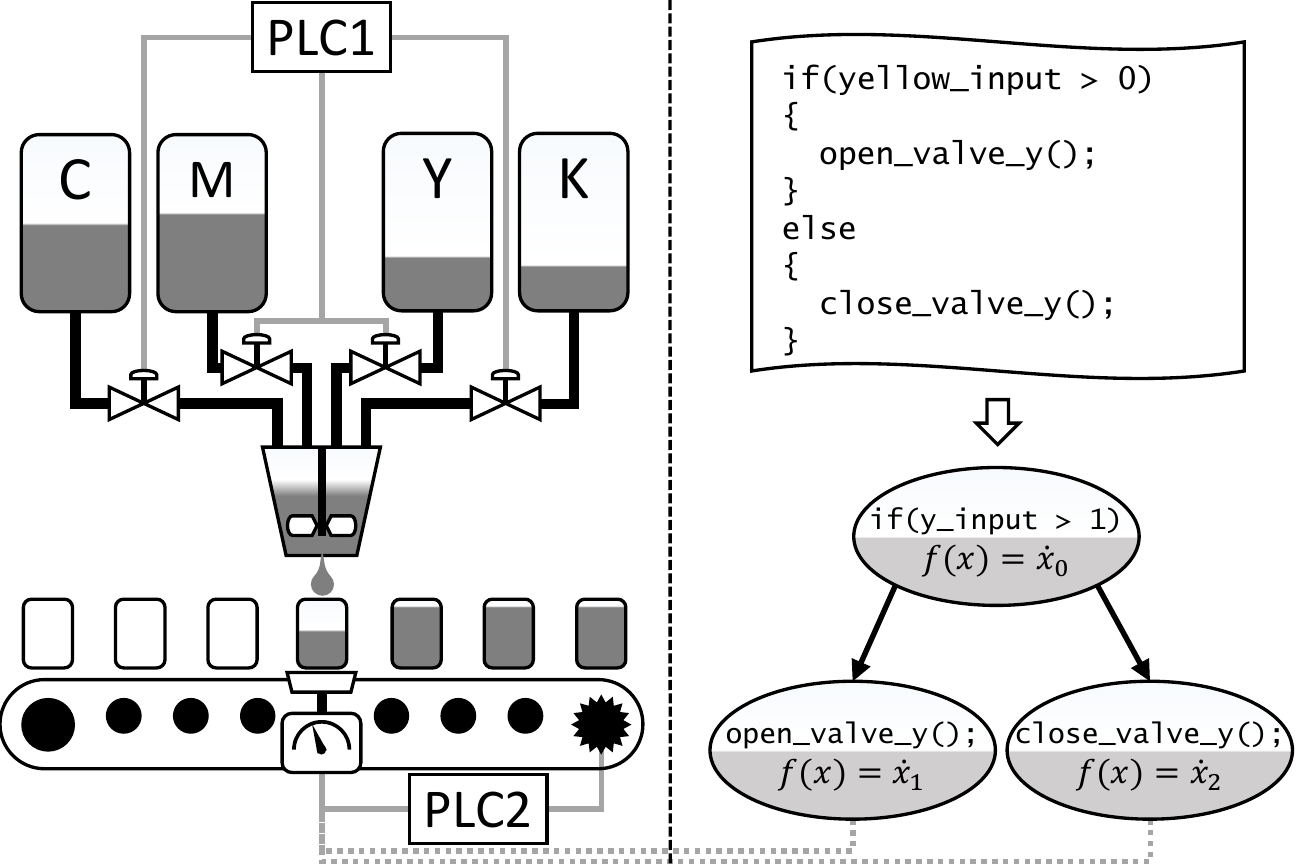}
	\caption{Simplified Industrial Control System, two PLCs control the mixing of four input colors--cyan (C), magenta (M), yellow (Y) and black (K for Key)--and filling into cans. PLC1's control actions on the valves of the individual color tanks will have an effect on the filling rate measured by PLC2.}
	\label{fig:ics}
\end{figure}

For instance, in \Cref{fig:ics} a simplified example of two PLCs controlling the mixing and filling of colors is shown.
Four input colors are mixed and filled into cans.
The input of each color is controlled by PLC1, which controls the respective valves.
PLC2 controls the conveyor belt, using a scale to determine when the current can is full and the next one has to be placed under the mixer.
The pseudo code and control-flow graph show the relation between the actions of PLC1 and the readings of PLC2, i.e., the operations of PLC1 determine the physical behavior observed by PLC2.
In such a hybrid system, the closing of the valve is a discrete event. The time required to fill a single can, on the other hand, will increase gradually (evolve continuously).


In the context of ICS, state estimation techniques have been leveraged to model physical dynamics for particular discrete events of the system~\cite{davis2015cyber,etigowni2016cpac}. 
However, the actuation of the associated controlling devices are typically abstracted to simplify the complexity of the model, neglecting the underlying control-flow behavior of any running programs. 
This simplification opens up these systems to motivated adversaries that exploit such abstractions to launch stealthy attacks~\cite{kangAdepuJacksonMathur}.

%% file: 3-systemmodel.tex
\section{Models and Assumptions}
\label{sec:model}

In this section, we present the system model and adversary model considered in this paper.

\subsection{System Model}
\label{sec:model:system}
We consider large distributed industrial control systems (ICS) with a centralized monitoring system (SCADA). 
This is the predominant system design~\cite{stouffer2011guide} for large scale industrial plants.
The ICS consists of networked controllers (\acsp{plc}) that jointly control a (complex) physical process, where the actions of the individual \acp{plc} are interdependent.
In particular, actuations initiated by one PLC effect the system state which will be represented in the sensor readings of other \acp{plc}.
This means that all \acp{plc} are indirectly connected with each other through the physical dynamics of the controlled physical system.\footnote{Note, the system does not need to be ``fully'' connected, i.e., the actuations of one PLC are not required to be observed by all other \acp{plc}. }

Each PLC is connected to its own local \red{array} of sensors and actuators.\footnote{\red{For simplicity we assume local sensors and actuators, however, remote I/O devices, i.e., networked sensors and actuators, can be modeled in our system as ``\acp{plc} without computations''.}} 
These sensors and actuators are directly interfacing with the physical system, and associated discrete sampled values are accessed by the \acp{plc}.

In addition to the indirect connection between \acp{plc}, all components of a distributed ICS are also connected explicitly. That said, all \acp{plc} are connected to each other and to SCADA over a computer network, e.g., Ethernet.
By means of the computer network, input and output data of all \acp{plc} are reported to the SCADA system, where data is recorded in a historian database and can be viewed by the operator.

\subsection{Adversary Model}
\label{sec:advmodel}

The adversary's goal is to cause misbehavior of the ICS while remaining undetected.
The behavior of the system refers to the actions that influence the physical process controlled by the ICS.
In particular, the adversary alters control commands sent to physical appliances (actuators) that can change the state of the physical process.
Passive attacks that do not alter the system behavior, e.g., attacks that ex-filtrate data, are out of scope in this work.

ICS usually have built-in safety functions that will be triggered by rapid changes of the system state or control commands that set system parameters far outside of the valid range.
Therefore, the adversary needs to make sure not to trigger these safety mechanisms.
Similarly, ICS are usually monitored by a human operator. 
The adversary has to make sure that her manipulations do not cause suspicion on the operator's side~\cite{garcia2017hey}, i.e., the attack needs to be stealthy.
This precludes na\"ive attacks like denial-of-service (DoS) on devices or the network.


\paragraph{Number of Compromised PLCs:}
The adversary can compromise one (or a small subset) of PLCs in a distributed ICS. 
The attacker in our adversary model has knowledge about the attacked system, so we have to assume a compromised PLC will report legitimate sensor values. Furthermore, we assume that the attacker is not able to compromise all PLCs in the ICS. We argue that this is a realistic assumption due to different reasons.
For instance, in a geographically distributed ICS, the adversary might have physical access to some PLCs in a remote station of the plant.
This is relevant if the attacker compromises PLCs via physical access, e.g., by updating the control logic via USB, replacing storage media like an SD-Memory card, or through debugging interfaces like JTAG~\cite{harvey}.
Furthermore, PLCs often have physical switches that deactivate the remote update functionality, i.e., an adversary has to have physical access to a PLC before being able to modify its software remotely.
Other reasons why an adversary cannot compromise all PLCs of a plant include systems which consist of heterogeneous PLCs, i.e., PLCs with different hardware or firmware versions, different models, or even from different vendors. 
If the adversary has knowledge about a vulnerability in one of the PLC variants, he can compromise these but not the other PLCs of the system.
Also, PLCs might be isolated in different network segments, exposing only a subset to a remote attacker.

We assume that the adversary has complete control over the compromised PLC, i.e., she can compromise the firmware and the control logic of the PLC.
The adversary can gain control over a PLC leveraging static or dynamic attack techniques. 
In a static attack, the adversary replaces the software (firmware or control logic) of a PLC, e.g., via a malicious software update.
Dynamic attacks are based on injecting new code at run-time, manipulating the behavior of existing code by means of return-oriented programming (ROP)~\cite{rop}, or data-oriented programming (DOP)~\cite{dop}.

\paragraph{\name:}
We assume that \name itself is not compromised. 
\red{\name executes on a separate system that is isolated from all other system components, including operator workstations and PLCs.}
\red{Run-time attacks that can compromise both a \ac{plc} and \name at the same time are particularly hard to find due to the architectural differences between them. \acp{plc} are typically ARM or MIPS based systems while \name is typically executed on an x86-based computer.} 
\red{Further, we assume that the \name system is hardened against cyber attacks using well-known defense mechanisms--e.g., CFI~\cite{abadi2005control}, or SoftBound~\cite{SoftBound} and CETS~\cite{CETS}--and protected against physical attacks, e.g., by being placed in a physically protected environment.} 
\red{Malicious updates of the \name system, e.g., through compromised operator workstations, can be countered with standard methods such as digital signatures and two factor authentication.\footnote{Such methods cannot be easily retrofitted to protect software updates of the \acp{plc} themselves due to resource limitations and legacy compliance requirements.}}


\paragraph{Network Attacks:}
For the sake of simplicity we consider network attacks out of scope.
We assume a secure, i.e., integrity protected and authenticated channel between the controllers and \name.
We discuss network attacks and defense mechanisms for settings without secure channels in \Cref{sec:network-attacks}.


%
%
%
%
%
%
  

%% file: 4-framework.tex
\section{Our Design}
\label{sec:design}

Before we describe our \name design and framework, we discuss important challenges that we had to tackle for \name.

\subsection{Challenges}
\label{sec:model:requ}

Important limitations in ICS stem from closed source, proprietary software.
Control software, firmware, and compilers are usually manufacturer specific and cannot be modified by the customer.
Thus, modification of the software running \emph{on} the PLC is not feasible as it would require cooperation of the manufacturer.
\red{The control logic, on the other hand, is usually developed by the plant operator, i.e., the operator has full access to its source code.}

Modifications of the PLC software also can lead to undesirable implications that will hinder adoption in practice.
Safety and reliability are paramount in ICS. Hence, all modifications that could impact them are unlikely to be adapted.
In particular, in systems that require safety certification modifications of the PLC software would void them, i.e., solutions that rely on the modification of control-components cannot be used in highly-sensitive environments.

\green{Finally, having a comprehensive view of the system is necessary to detect the various attacks mentioned before. Although a single PLC may have access to monitor the values of other sensors/actuators of the ICS, the maintenance of state estimation of the physical processes will incur significant overhead in the PLC scan cycle. Even if a PLC had the memory resources for such state estimation, the computation of the state estimation may cause a violation of the real-time constraints.}


\subsection{\name Design}

\begin{figure*}[!t]
	\centering
\includegraphics[trim = 0mm 0mm 0mm 10mm, clip, width=.85\textwidth]{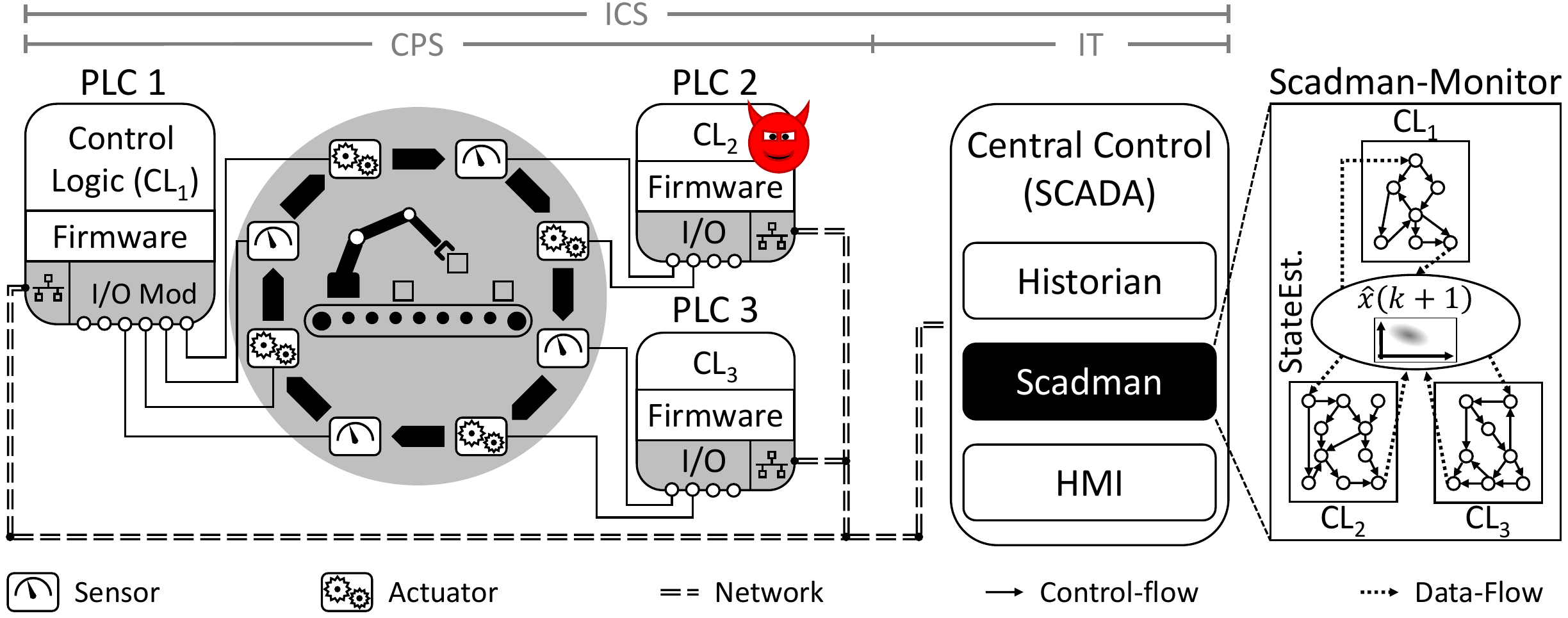}
	\caption{\name system overview and architecture. The central control (SCADA) is extended with a component called \masterplc that monitors the behavior of the distributed ICS and can detect compromised controllers in the system.}
	\label{fig:design}
\end{figure*}

The goal of \name is to ensure the correct behavior of a distributed industrial control system (ICS).
The correct behavior can be violated by different types of attacks, as discussed before in \Cref{sec:advmodel}.
As a result, \name must provide a general mechanism that can counter all possible attacks that result in an incorrect control behavior of the system. 
Control behavior includes any action taken by any of the \acp{plc} that modifies the overall system state. 
We consider the control behavior as correct, if it fits the behavior intended by the system operator.
An attack can result in incorrect control behavior, when the attacker makes one of the components perform a different action than was intended. 
For example, a \ac{plc} is \red{supposed} to close a valve when a certain threshold is reached. The attacker then forces the \ac{plc} to keep the valve open, contrary to the original programming of the \ac{plc}.
\name ensures the \emph{control behavior integrity} (CBI) of an ICS. A violation of CBI is a deviation from the intended behavior of any of the \acp{plc} within the \ac{ics}. 

The system can also deviate from the intended state for other reasons like faults, e.g., a faulty sensor reporting incorrect values. \name can detect these situations, allowing the operator to repair the system.

%

\Cref{fig:design} shows the concept of \name.
In a distributed ICS, multiple \acp{plc} interact independently with a physical process.
However, the actions of one \ac{plc} influence the overall state of the physical system. This is reflected in the sensor readings of other \acp{plc}. 
We exploit this interdependency to detect the misbehavior of a compromised PLC. For example, a compromised \ac{plc} cannot stealthily open a valve because a second trustworthy \ac{plc}\red{, which is not under attacker control,} would measure the change in inflow. This discrepancy between expected sensor readings and the actual system state is used by \name to detect deviations in the control behavior.

\paragraph{\masterplc:} All PLCs report their actuation commands and sensor readings to a central entity, which we call \masterplc. 
The \masterplc is a program that interacts with a simulated physical system and allows \name to calculate the expected state of the overall \ac{ics}.
Note that centrally reporting and logging all operations is very common in ICS~\cite{stouffer2011guide}, e.g., for reporting to HMI components. 
Based on the retrieved data, the \masterplc will subsequently check whether any of the \acp{plc} have been deviating from the intended behavior\red{, i.e., that all \acp{plc} have been following a small set of valid control flow paths given the current system state}.
This check requires two components of the \masterplc. (1)~A consolidated control logic code of \emph{all} PLCs, and (2)~a model of the physical process allowing \masterplc to determine the interdependencies of the PLCs' inputs and outputs\red{, shown on the right in \Cref{fig:design}}.

\name generates the consolidated control logic which combines the control logic of \emph{all} PLCs into a single large program that represents the entire control actions of the ICS.
This code is executed on the \masterplc to determine valid actions of the \acp{plc}.
Based on the current state of the overall system and the model of the system's physical state, \name dynamically derives the legitimate control-flow paths through the \ac{plc} code in a physics-aware manner.
This means, \name does not accept all possible, benign control-flow paths in the control logic's control-flow graph as valid, as is the case with CFI~\cite{ABE+05,ABE+09}, but only those that are valid at any given time in the current state of the \ac{cps}.
This approach limits the set of allowed control-flow paths and thus the adversary's actions.

The physical process model allows the \masterplc to estimate the influence of control commands sent by \acp{plc} on the expected sensor readings. 
As the adversary cannot influence the physical model of the system (the laws of physics cannot be altered), an inconsistency between actuation commands and sensor readings implies that  either the \ac{plc} controlling the actuation or the \ac{plc} controlling the sensors must behave incorrectly\red{, i.e., issue wrong actuation commands or report forged sensor readings.}
\name can tolerate imprecise and incomplete models. The model quality largely determines the detection precision. However, an imprecise model, noisy sensors, and other factors impacting the state estimation are handled by \name as described below and in more details in \Cref{sec:multi-exec}.

\subsection{\name Framework}

\begin{figure}[th]
	\centering
	\includegraphics[width=.9\linewidth]{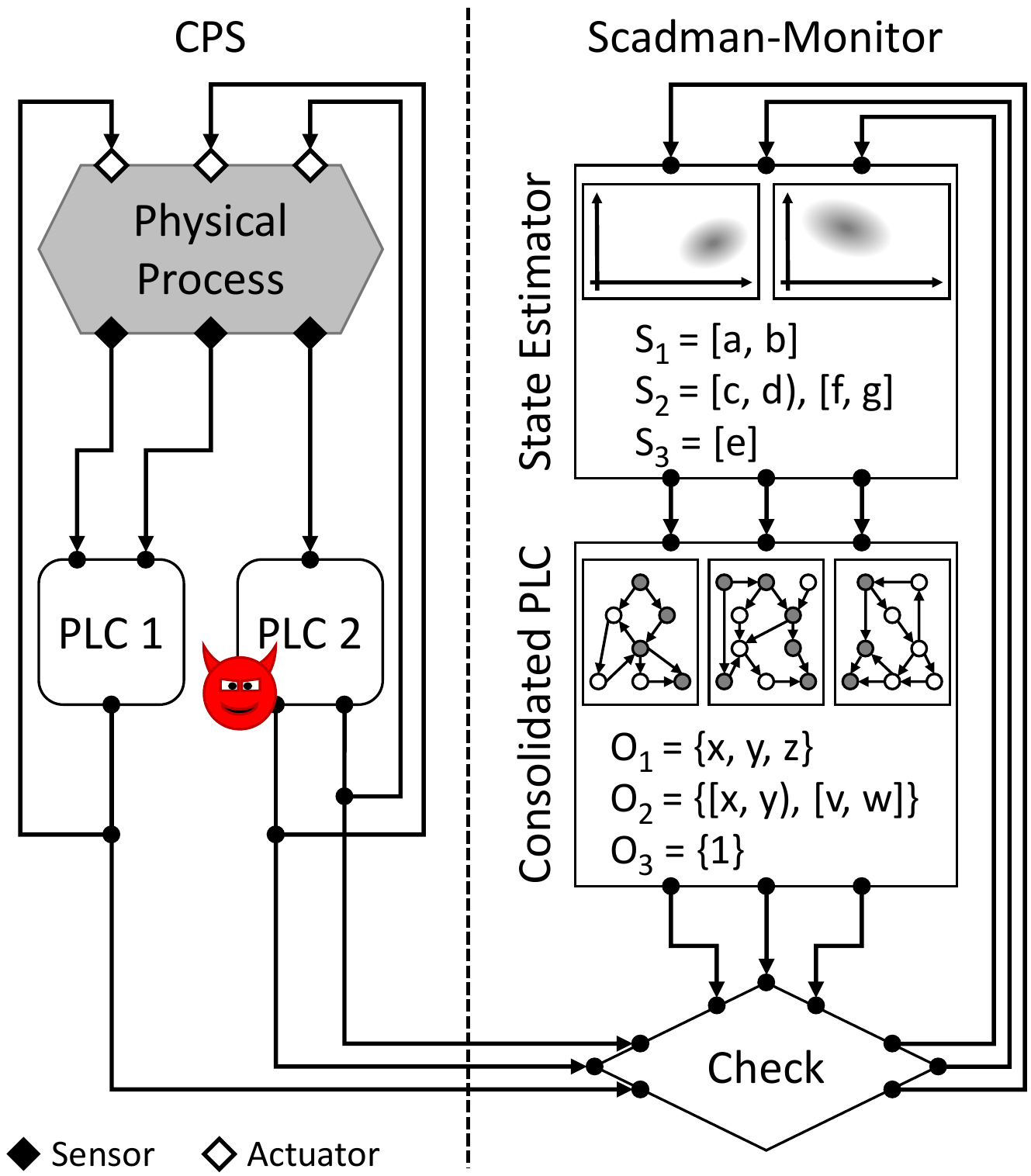}
	\caption{\name scan cycle. For each interaction of a PLC with the physical process, the \masterplc performs the corresponding operations using the state estimated based on the system's past events. Afterwards, consistency between both sides \red{(\ac{cps} and \masterplc)} is checked, and an alarm is raised when a deviation is detected.}
	\label{fig:cycle}
\end{figure}

Our \name framework leverages a compiler-based approach to automatically generate the \emph{consolidated PLC} code and connect it to the physical \emph{state estimator}.
Both are executed in \masterplc, as shown in \Cref{fig:cycle}.

\paragraph{Physical State Estimation:}
The state estimator uses physical models of the physical processes that are controlled by the \acp{plc} of the \ac{ics}.
The state estimator simulates the evolution of the physical system.
Based on the current state of the physical systems and actuation inputs to the system the state estimator determines the following state of the system.

Physical systems usually evolve continuously, however, for \name the system state at the sampling point is relevant, i.e., at the points in time when a \ac{plc} reads the system state using its sensors.
Models of the physical processes in \ac{ics} are usually known by the operator of the plant. 
Additionally, \green{several recent works have} developed methods to extract and generate such models, which can be used with \name~\cite{davis2015cyber,etigowni2016cpac}.

\paragraph{\masterplc Operation:}
\name runs in parallel with the \ac{cps} (on the left in \Cref{fig:cycle}) and compares the sensor readings and actuation commands it receives over the network from all \acp{plc} to a set of valid state determined by \masterplc--allowing for the validation of the behavior of the \ac{cps} (i.e., distributed \ac{ics}). 

\masterplc works in iterations, similar to the scan cycle based operations of \acp{plc}.
During each iteration the current system state and actuation commands are fed into the state estimator of \masterplc, in parallel the physical process in the \ac{cps} evolves, based on the current system state and past actuation commands.
The state estimator calculates the state space into which the system should have evolved based on its history, in \Cref{fig:cycle} the state variables $S_{1} \dots S_{3}$ have been predicted to lay within some given interval of possible values. 
The fuzziness of the state space can be, for instance, due to impressions of the physical model used in state estimator or due to (small) errors in the input value.
However, \name can tolerate these impressions. 

The actual physical state in the \ac{cps} is read by the \acp{plc}’s sensors and processed by them.
In parallel, \masterplc executes the consolidated PLC using the estimated state as input.
Since the input state can be fuzzy the execution of \masterplc has to account for it.
By using a technique called \emph{error-margin multi-execution} the execution is performed over the entire range of possible input values (see \Cref{sec:multi-exec} for details).
By executing--given the current system state context--\emph{all} valid control-flow paths in the consolidated \ac{plc} \masterplc determines the set of possible outputs.
Again, the outputs ($O_{1} \dots O_{3}$ in \Cref{fig:cycle}) are represented as intervals or sets of allowed values.

When \masterplc receives the sensor reading and actuation command of the \acp{plc} from the \ac{cps} it performs the consistency check.
If all \acp{plc} were executing correctly--and all sensors and actuators operated correctly--the reported values must be a subset of the outputs determined by \masterplc.
Any deviation indicates an inconsistency within the behavior of the \ac{cps} and \name rises an alarm.

If the system was found to be correct the system state reported by the \acp{plc} is accepted as the current system state and serves as input for the next iteration of \masterplc.
This is necessary to prevent the system state calculated by the state estimator to gradually deviate from the actual system state.\footnote{To detect and counter slow evolving attacks \name can be adapted to simulate the system state over multiple iterations at the cost of added impression in the state estimation (see \Cref{sec:discussion}).}

%% file: 5-llvm-instrumentation.tex
\section{\name Implementation}
\label{sec:impl}
\name introduces the \masterplc process, which is responsible for receiving all sensor values and actuations, and validating them using the consolidated PLC code and state estimation. We propose a generic approach to build the \masterplc. 
We will first discuss how \name consolidates the control programs for a distributed PLC network along with the necessary assumptions for timing and functional correctness. We then demonstrate how the consolidated code will be compiled and instrumented into an executable that can receive the state of the ICS network, e.g., the state of the sensors, as input for each scan cycle and update the estimated state of the system. We also introduce a novel approach, so-called error-margin multi-execution that allows \name to account for \red{cases where} the model of the physical system deviates from the real system. Finally, we describe how these components of \name can be combined with the physical state estimation to detect any compromised components in the ICS.

\begin{figure}[!thb]
%
\begin{minipage}{.47\linewidth}
\centering
\begin{lstlisting}[
    frame=single,
    basicstyle=\tiny
]
PROGRAM plc1
  VAR_INPUT
    YellowAmount : REAL;
  END_VAR
  VAR_IN_OUT
    YellowValve : BOOL;
  END_VAR
  IF (YellowAmount > 0) THEN
    YellowValve := 1;
  ELSE
    YellowValve := 0;
  END_IF; 
END_PROGRAM
CONFIGURATION Config0
  RESOURCE Res0 ON PLC
    TASK Main(INTERVAL := T#1s,
              PRIORITY := 0);
    PROGRAM Inst0 WITH Main : plc1;
  END_RESOURCE
END_CONFIGURATION
\end{lstlisting}
\vspace{-1px}
\begin{lstlisting}[
    frame=single,
    basicstyle=\tiny
]
PROGRAM plc2
  VAR_INPUT
    CanWeight : REAL;
    YellowValve :  BOOL;
  END_VAR
  VAR_IN_OUT
    ConveyorMove : BOOL;
  END_VAR 
  IF (CanWeight > 100.0 
      AND NOT(YellowValve)) THEN
    ConveyorMove := 1;
  ELSE 
  	ConveyorMove := 0;
  END_IF;
END_PROGRAM
CONFIGURATION Config2
  RESOURCE Res0 ON PLC
    TASK Main(INTERVAL := T#1s,
              PRIORITY := 0);
    PROGRAM Inst0 WITH Main : plc2;
  END_RESOURCE
END_CONFIGURATION
\end{lstlisting}
\end{minipage}
\hfill
\begin{minipage}{.47\linewidth}
\begin{lstlisting}[
    frame=single,
    basicstyle=\tiny
]
(* Master PLC Code *)
PROGRAM master

  (* combined  variables *)
  
  VAR_INPUT
    YellowAmount : REAL;
    CanWeight : REAL;
  END_VAR
  
  VAR_IN_OUT
    YellowValve : BOOL;
    ConveyorMove : BOOL;
  END_VAR

  (* plc1 code*)
  
  IF (YellowAmount > 0) THEN
    YellowValve := 1;
  ELSE
    YellowValve := 0;
  END_IF;

  (*plc2 code*)
  
  IF (CanWeight > 100.0 
      AND NOT(Filling)) THEN
    ConveyorMove := 1;
  ELSE 
  	ConveyorMove := 0;
  END_IF;
END_PROGRAM

(* master  configuration*)
CONFIGURATION MasterConfig
  RESOURCE Res0 ON PLC
    TASK Main(INTERVAL := T#1s,
              PRIORITY := 0);
    PROGRAM Inst0 WITH Main: master;
  END_RESOURCE
END_CONFIGURATION
\end{lstlisting}
\end{minipage}

\hfill
\caption{Code consolidation for two PLCs with respect to a single process of the paint mixing plant in \Cref{fig:ics}. The left 2 programs, \texttt{plc1} and \texttt{plc2} are merged into a \texttt{master} PLC code.}
\label{fig:simple-consolidation}
\end{figure}
\noindent\textbf{Example.} 
%
%
%
%
%
For the purpose of clarity, we will provide a simplified representation for a single process control for two PLCs from the system in \Cref{fig:ics}. \Cref{fig:simple-consolidation} shows two PLC programs, \texttt{plc1} and \texttt{plc2} that are consolidated by \name into a single PLC representation, \texttt{master}. \plc{1} is responsible for controlling a valve, \texttt{YellowValve}, associated with the yellow color dispenser. The valve will open if an input amount is greater than 0. \plc{2} is responsible for moving the conveyor belt if the current can is full and if the \texttt{YellowValve} is not open. Descriptive variable names have been used in the code. This example will be used to explain each component of the implementation.



\subsection{PLC-Code Consolidation}
The premise of generating a \masterplc representation is to first merge the control program code of all the ICS PLC's into a single PLC program representation. 
This consolidated representation is necessary for two reasons. First, in order to monitor the distributed processes, we need access to all of the system parameters in order to successfully simulate the physical model of the overall system, i.e., we cannot simulate the physics of a process with partial sensor data. 
In theory, the consolidation would not be necessary for a subset of the PLCs that do not have any cyber-physical interdependencies. However, these dependencies are difficult to derive. As such, \name automatically generates models that incorporate these cyber-physical interdependencies as long as the distributed system conforms to the assumptions required to ensure functional and timing correctness, which are discussed at the end of this subsection.

In this paper, we consider PLC control programs that conform to the IEC 61131 standard~\cite{john2010iec}. According to the standard, programs are typically composed of three types of programming organisation units (POUs): programs, functions, and function blocks. A  \textit{program} is the ``main program'' of the PLC that includes I/O assignments, variable definitions, and access paths. A \textit{function} is a programming block that returns a value given input and output variables in a similar vein to function definitions for other procedural programming languages such as C. A \textit{function block} is a data structure that has the same functionality as a function but retains the associated values in memory across executions. As such, the code consolidation process will append all of the function and function block definitions and merging the main PLC program of each PLC. This allows us to retrieve the state of all sensors and actuators of the ICS, feed the values through this consolidated representation, and observe how the actuators are updated. \Cref{fig:simple-consolidation} illustrates how the main programs of two PLC programs will be merged.

However, it is common practice to define different components for a single PLC control program using different programming languages.  The IEC 61131 standard enumerates five programming languages: (1)~ladder diagrams (LD) -- a graphical programming language to design logic circuits, (2)~sequential function charts (SFC) -- another graphical programming language to define sequential state operations, (3)~function block diagrams (FBD) -- a graphical representation of function blocks, (4)~instruction lists (IL) -- an assembly-like textual programming language, and (5)~structured text (ST) -- a textual programming language similar to Pascal. The heterogeneity of a PLC program significantly increases the complexity of any form of static code analysis as compilation and simulation rules would have to be defined for each language. As such, \name first converts all programs to a single programming language representation. Previous works have formally proven that the structured text (ST) programming language can be used to represent the other four languages~\cite{darvasgeneric} and therefore serves as our base programming language for the \masterplc.

We will now discuss the correctness of our consolidation process and the necessary assumptions.

\noindent\textbf{Timing correctness.} The correctness of consolidating the control logic of all \acp{plc} depends on the required sampling time of the ICS. \ac{plc} tasks can be executed either continuously or periodically for some interval. For a distributed network of $N$ \acp{plc} that are configured to run programs at varying sample times, $T_{sample}(i)$ for a $\mathrm{PLC}_{i}$, the consolidation is valid if and only if the sum of the execution times, $T_{execution}(i)$ of all controller programs is less than the smallest task interval, i.e., 
\begin{center}
 \vspace{-0.08in}
$\sum_{i=1}^{N}T_{execution}(i) < \min_{\forall{i} \in N}T_{sample}(i)$.
 \vspace{-0.08in}
\end{center}

For continuously executing \ac{plc} configurations--i.e., event-driven control--the cumulative scanning time must be less than the shortest duration time of an input or an output signal~\cite{moon1994modeling}. 
The continuous scan cycle time of \acp{plc} range from microseconds to tenths of a second. However, because \name is implemented on a standard computer with substantially more computing power than typical \acp{plc}, the ``scan cycle'' for \name's consolidated \ac{plc} code is much faster and, hence, the only bottleneck is the sampling time over network communication. In the system shown in ~\Cref{fig:simple-consolidation}, the programs \texttt{plc1} and \texttt{plc2} are shown to have the same execution interval timing, which is reflected in the consolidated \texttt{master} program.

With respect to clock drift between \acp{plc}, we assume that the design of the overall ICS accounts for clock drift as such a hindrance would be a pre-existing condition.

\noindent\textbf{Functional correctness.} We also consider the functional correctness of combining multiple control logic programs sequentially into a single control logic program. A \ac{plc}'s scan cycle can be abstracted into three components: the scanning of the inputs, the propagation of the inputs through a logic circuit, and the updating of all associated outputs at the end of the scan cycle. The \masterplc program combines the scanning of inputs and updating of outputs for all \ac{plc} programs as these actions are atomic in nature. Previous work has shown that separate processes update the values of inputs/outputs to/from memory independent of the scan cycle process~\cite{garcia2017hey}. 

Because the process of propagating the inputs through a logic circuit to update the associated outputs is parallel in nature across \acp{plc}, we can inductively claim the correctness of merging based on the timing correctness of our assumptions. Furthermore, the ordering of the merging process is arbitrary as any unsatisfied dependencies or race conditions amongst \acp{plc} would be a pre-existing nuisance in the design of the system. For instance, for the system in \Cref{fig:simple-consolidation}, the ordering of \texttt{plc1} and \texttt{plc2} is arbitrary in the context of the \texttt{master} program. If there was a race condition and/or ordering dependency where both programs were writing to the actuator \texttt{YellowValve}, this would be a flaw by design of the overall ICS. 

\subsection{Compilation and Instrumentation}
\begin{figure*}[th]
  \centering
  \includegraphics[width=0.8\linewidth]{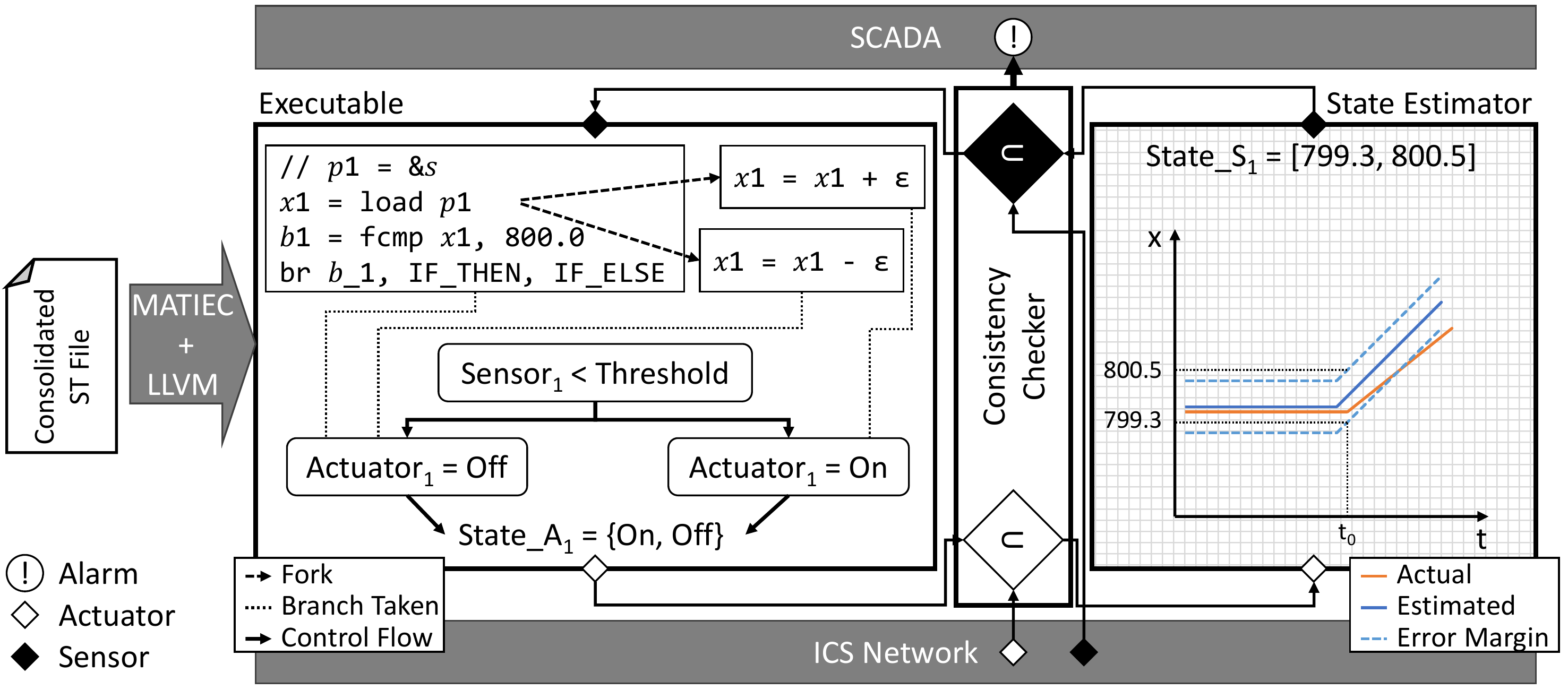}
  \caption{\name implementation overview. The consolidated PLC code is compiled in combination with MATIEC and LLVM to an executable. The ICS network feeds sensor values into \masterplc to simulate the PLC scan cycle and check for consistency. The updated outputs are fed from the executable to the state estimation as well. Any deviations in the expected behavior is alerted to the SCADA monitor. }
  \label{fig:implementation}
\end{figure*}
To compute the updated values of actuators at the end of a scan cycle, we need to execute the consolidated PLC code given the state of the sensors. We integrate the consolidated control code into the \masterplc and record the actuations, performed by the control code. \Cref{fig:implementation} provides an overview of the implementation and shows how the consolidated code is incorporated into the \masterplc.
We use \red{our extended} MATIEC compiler\footnote{\url{https://github.com/thiagoralves/OpenPLC_v2/}} to compile the consolidated PLC code to C code. We modified the MATIEC compiler to automatically generate functions that allow easy access and modification of the internal state of the generated C code. This is used by the \masterplc process to simulate access to sensor values and actuators. 

As a second step, we compile the C code into an intermediate representation using the LLVM~\cite{LLVM:CGO04} compiler framework. Operating on the LLVM intermediate code allows us to perform analysis and instrumentation without having to analyze structured text or C code directly. We perform instrumentation on the generated LLVM intermediate code to introduce an 
execution mode that draws from the ideas of symbolic execution, abstract interpretation and interval arithmetic. We call this execution mode \textit{error-margin multi-execution}. We use this execution mode to reduce the number of false positives by introducing \red{the notion of an error-margin} to accessed sensor values. We discuss details of this approach in the following subsection.

The final step is to produce a runnable executable. We compile the instrumented PLC code to native code and link it with \red{our} support library, providing various utility functions. The C code generated by the MATIEC compiler is intended to be linked to a userspace driver that implements hardware access. Instead we link the generated code with our framework such that all hardware accesses are intercepted and forwarded to the state estimation and attack detection.
%

\subsection{Error-margin Multi-execution}
\label{sec:multi-exec}
The model of a physical system may deviate from the real system due to various reasons.
For instance, a physical processes might evolve slower or faster than expected in the model. 
These minor differences between the model-based estimated state and the real system state can lead to inconsistencies that \name would incorrectly report as an attack. 
These false positives can occur, for instance, if the PLCs perform an actuation depending on whether a sensor value is above or below a threshold, as shown in the example in \Cref{fig:implementation}.
In the real system, the sensor might be above the threshold, while in the simulated physical system, the sensor is still below the threshold. 
In this case, the \masterplc performs an actuation, while the real PLC does not, or vice versa. For instance, this can happen when the weight of a can increases at a slightly higher rate than estimated by the physical system model. 
The reason for this can be a valve, which controls the inflow of a color into a can and may not necessarily close within one scan cycle. 
In our system model, the actuations are assumed to be immediate, i.e., the valve will be closed and our system model will reflect an inflow rate of 0. 
In reality it may only be partially closed with a nonzero inflow rate. 
These slight deviations will be propagated to the associated physical model.

To tackle this problem and reduce false positives, we check whether the \masterplc behaves differently in terms of actuation assuming an error in the sensor readings. We introduce error-margin multi-execution to detect differences in actuation. First, we define error-margins for sensors. Second, we detect whether a \ac{plc} performs different actions, when executed with an error applied to the sensor value. A difference in the performed actions, are only observed when the \ac{plc} is taking a different control-flow through the program execution. Therefore, we need to detect whether the control-flow of the \ac{plc} depends on a sensor value (cf. code snippet in \Cref{fig:implementation}). 

We define an error-margin, \(\pm\epsilon\), for each of the sensors. 
We then check whether the \masterplc performs different actuations when applying \(\pm\epsilon\) to the sensor reading, which we denote as \(s\).
Using interval arithmetic, one can propagate the error-margin through the executed program. 
However, if a branching condition depends on the sensor value, possibly two branches must be executed if the decision is inconclusive. 
For example, the branching condition is \(( s < N )\), then the execution could take both branches if \(s + \epsilon \geq N\) and \(s - \epsilon < N\). 
Therefore, we need to execute multiple paths through the control program. 
Symbolic execution would allow us to use symbolic sensor values and constrain them into the error-margin and execute multiple paths at once. 
However, current symbolic execution engines have known limitations when it comes to solving constraints for floating point operations~\cite{Liew2017-yh}. Typically sensor values are represented as floating point types. 
To overcome this limitation we introduce multi-execution that operates solely on concrete floating point values within the error-margin applied and can execute multiple branches in parallel.

We integrate this error application into the consolidated PLC code simulation at the LLVM level. 
Whenever a conditional branch instruction depends on a sensor value $s$, we introduce instrumentation that forks the execution of the PLC code. 
In one fork we continue without an error, so $s' = s$. 
In the second and third fork we continue with the upper bound of the error-margin $s' = s + \epsilon$ and the lower-bound $s' = s - \epsilon$, respectively. 
Using \emph{only} the upper and lower bound of the error interval $[s - \epsilon, s + \epsilon]$ is not sufficient.
To be able to evaluate equality comparisons we need to also continue one fork of the code on $s$ (without applying any error).
At the end of the scan cycle we merge all forks and continue without any error applied in the next scan cycle. 
We create \(\mathcal{O}(3^{\#sensors})\) forks per scan cycle. 
While this is a significant overhead in the worst case, \red{our evaluation, however, shows the practicality of this approach.} We can use several optimizations in practice to reduce the number of concurrent forks. 
For example, if two forks take the same control-flow path, we can stop executing one of the two forks. Most basic blocks have only two outgoing edges, therefore we can usually kill one of the forks directly after they have taken the branch. 
In fact, we do not need to use the multi-execution approach until we detect a discrepancy in the actuations. We can then selectively re-execute only the violating scan cycle in multi-execution mode to get a more accurate result on the actuation. 



Instead of producing one value for an actuator we now get a set of values for each of the actuators. 
If the actuation of the real system is not in the set which is reported by the consolidated PLC code we detected an inconsistency which is beyond the errors-margins and report an attack.
Incorporating state estimation errors allows \name to minimize false positives that would arise with slight errors in the model of the physical system.

To detect whether a branch condition depends on a sensor value, we perform backwards data-flow analysis, starting from the condition of the branch. 
We use the \ac{ssa} of LLVM intermediate code to perform intra-procedural data-flow analysis. 
Inter-procedural analysis is not implemented in our current prototype as the code generated by MATIEC does not require inter-procedural data-flow analysis for most sensors. 
We search the resulting data-flow graph for load instructions that read sensor values. Because the load instruction is contained in the backwards data-flow graph, we know it will affect the branching condition and must be instrumented to incorporate the check for forking the process based on the given error-margins.

Our error-margin multi-execution introduces some imprecision into the system. 
An attacker might try to exploit this imprecision to evade detection, we discuss this scenario in more detail in \Cref{sec:security}.

\subsection{Attack Detection}

To detect attacks, the \masterplc performs two steps, where the results from the $n$-th scan cycle are used to predict and verify the $n+1$-th scan cycle of the system. 
First the \masterplc compares sensor values received in scan cycle $n$ with the sensor values estimated based on the inputs from scan cycle $n-1$. 
When the received senor values are verified, i.e., fall within the set of predicted values, the \masterplc uses them as input to execute the consolidated PLC code.
This results in a set of acceptable actuation operations for scan cycle $n$.
\masterplc compares this set against actuation operations reported by the real PLCs.
If the actuations of the real PLCs are verified correctly they serve as input to the state estimator of the physical system, which will predict the sensor values for the next scan cycle $n+1$.

\paragraph{Incomplete data.}
\name can also be used on system that cannot provide complete data or which involve (sub)process for which no accurate state estimation is possible.
This can be due to various reasons, e.g., if a sensor state depends on human interactions with the system the \masterplc cannot predict the state of that sensor in a meaningful way.
However, the state of other sensors and actuators of the system that are not directly influenced by such external influence can still be validated by \name.


%% file: 6-security.tex
\section{Security Considerations}
\label{sec:security}

In this section we consider different kind of attacks in the context of ICS and how \name can detect them.
According to our adversary model (cf.~\Cref{sec:model}) we consider a subset of PLCs to be compromised, i.e., $k$ out of $n$ PLCs are compromised, where $k < n$.

Afterwards we will discuss attacks scenarios that go beyond our adversary model and show that \name is valuable in these scenarios as well.

\subsection{k-out-of-n Compromised PLCs}
As discussed before, for various reasons the attack might have compromised a subset of PLCs in an ICS.
The adversary aims to act stealthy, hence, we assume the adversary lets the compromised PLCs report sensor readings and actuation commands that meet the expectations of the operator as well as \name.
However, the reported values will \emph{not} match the expected values relative to the values reported by the non-compromised PLCs.

In particular, as long as one PLC that is physically interconnected with the compromised PLCs reports correct values, a discrepancy will emerge.
\name will detect this discrepancy and will raise an alarm.
While \name will not be able to identify which PLCs are compromised, it can still warn the operator, who can then start an in-depth investigation on the system.
In the next section we evaluated \name on a large set of ICS attacks implemented for a real-world industrial control network (SWaT~\cite{swatDataset}) and show that it can detect all of these attacks.

\paragraph{Slow evolving attacks.}
\name is based on a closed loop approach where, for each scan cycle, the system is analyzed for anomalies. The system continues when no anomaly is detected.
By continuing, the current state of the system is accepted as benign and serves as the basis for estimating the system's future state. 
An adversary could try to exploit this scenario by \emph{slowly} pushing the system towards a false state.
On each iteration, the adversary would manipulate the system within the error margins of \name.
However, ICS are usually designed to include safety measures programmed into the PLCs that prevent the system from being steered to an unsafe state.
While the attack can slowly modify the system within the safety boundaries of the system without being detected, the system cannot be pushed to an unsafe state. \name would detect any deviations in the control flow path of the PLC, e.g., if an adversary pushes the system outside of the safety boundaries enforced by a safety check within the control flow of the original PLC program.
The best the adversary can do is to leverage the simulation error margin used by \name to get the system slightly outside of its safety boundaries.
However, the safety boundaries are chosen such that the system remains safe even in the presence of small errors, e.g., due to sensor measurement noise.

\red{We discuss additional counter measures in \Cref{sec:discussion}}

\subsection{All PLCs Compromised}
\name provides security based on the assumption that physical interdependencies of controllers (PLCs) enable the detection of misbehavior.
In the simple case that the entire system is controlled by a single PLC, an adversary would control all of the data (e.g., sensor readings) available to \name if this PLC is compromised.
Hence, an intelligent adversary can provide a consistent view of the system~\cite{garcia2017hey} towards \name and remain undetected.

For distributed ICS, the adversary needs to control \emph{all} PLCs to provide a coherent view of \emph{all} actuation commands and \emph{all} sensor readings reported to \masterplc.
This means the adversary has to simulate the expected behavior of the \emph{entire} system and synchronizes the actions of all PLCs.
While this might be feasible for very simple and static ICS, the attacker's limited resources (PLCs have limited computation power and memory) significantly aggravate the complexity of stealthy attacks for dynamic ICS.


\subsection{Network Attacks}
\label{sec:network-attacks}
In this work we assume a secure channel between \masterplc and the \acsp{plc}, i.e., an adversary cannot launch network attacks by impersonating other devices.
However, some legacy systems do not provide secure network channels, in which cases the adversary might try to overcome \name by manipulating network packages.

The adversary can either try to manipulate or suppress network packages of other PLCs, i.e., PLCs not controlled by the adversary that would reveal the adversary's behavior manipulations. 
This is not possible in commonly used switched networks, i.e., network packages of an un-compromised PLC will never be routed to a compromised PLC but directly to \masterplc.

The second option for the adversary is to impersonate another PLC, i.e., by sending packages to \masterplc pretending to originate from an uncompromised PLC, e.g., by modifying the source IP address of a package.
However, as discussed before, the adversary cannot suppress packages sent by the benign PLC, hence, \masterplc will receive both types of packages: those with benign sensor reading and those with manipulated values reported by the adversary.
This mixture of input values will lead to inconsistencies, which will trigger a security alarm of \name.

%% file: 7-evaluations.tex
\section{Evaluations}
\label{sec:eval}


In this section, we provide an overview of our experimental evaluation of \name. We first introduce the real-world industrial control network 
that we used for our evaluation. We then discuss how \name implements code consolidation for the proprietary PLCs used \red{in our evaluation}. Afterwards we describe the choice of physical state estimation equations used for our attack detection. Finally, we evaluate \name against a set of attacks that were enumerated by previous works.

\subsection{Evaluation Environment and Dataset}
The study reported here was conducted on data from a real distributed industrial control system~\cite{swatDataset} as shown below. 

\begin{figure}[!htb]
\centering
\includegraphics[width=.9\linewidth]{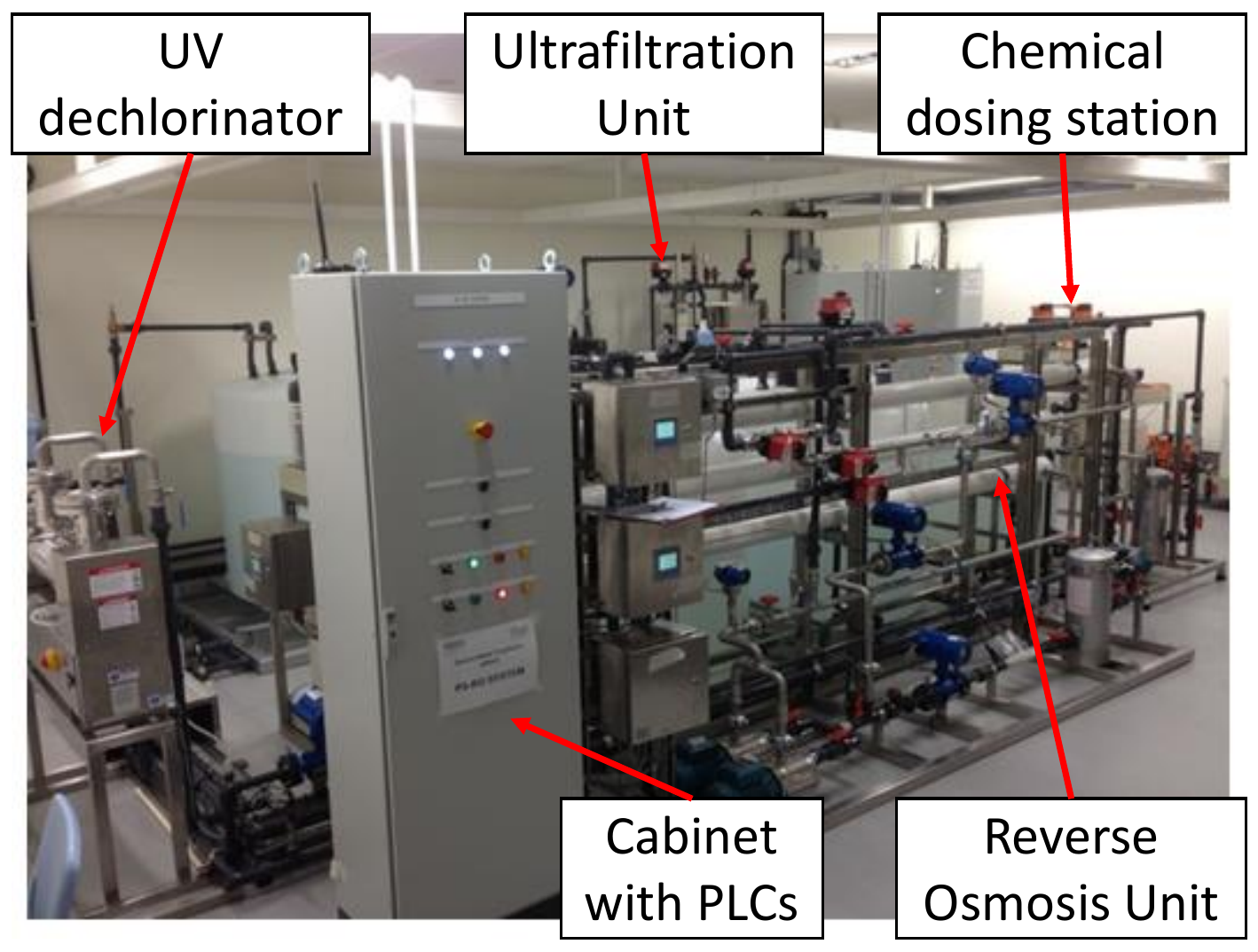}
\caption{Sub-processes in the industrial water treatment control network}. 
\label{fig:swatPhoto}
\end{figure}


The network is a 6-stage water treatment plant that produces 5~gallons/minute of treated water.  The plant can operate non-stop  24/7 in fully autonomous mode.
The sub-process of each stage is controlled by an individual PLC (cf. \Cref{fig:swatPhoto}.
In total, the plant contains 68 sensors and actuators; some actuators serve as standbys and are intended to be used only when the primary actuator fails.

\noindent {ICS operation}:  Operation of the plant is initiated by an operator at the SCADA workstation and,  when needed,  can be controlled. State information can be viewed at the workstation or at the HMI, and is recorded in the historian.

\noindent {Plant supervision and control}: A Supervisory Control and Data Acquisition (SCADA) workstation is located in the plant control room. Data or control access to nearly all plant components is available via this workstation.  A plant operator can view process state and set process parameters via the workstation.   A historian is available for recording process state as well as network packet flows at preset time intervals.

\begin{figure} [!htb]
\centering
\includegraphics[width=\linewidth]{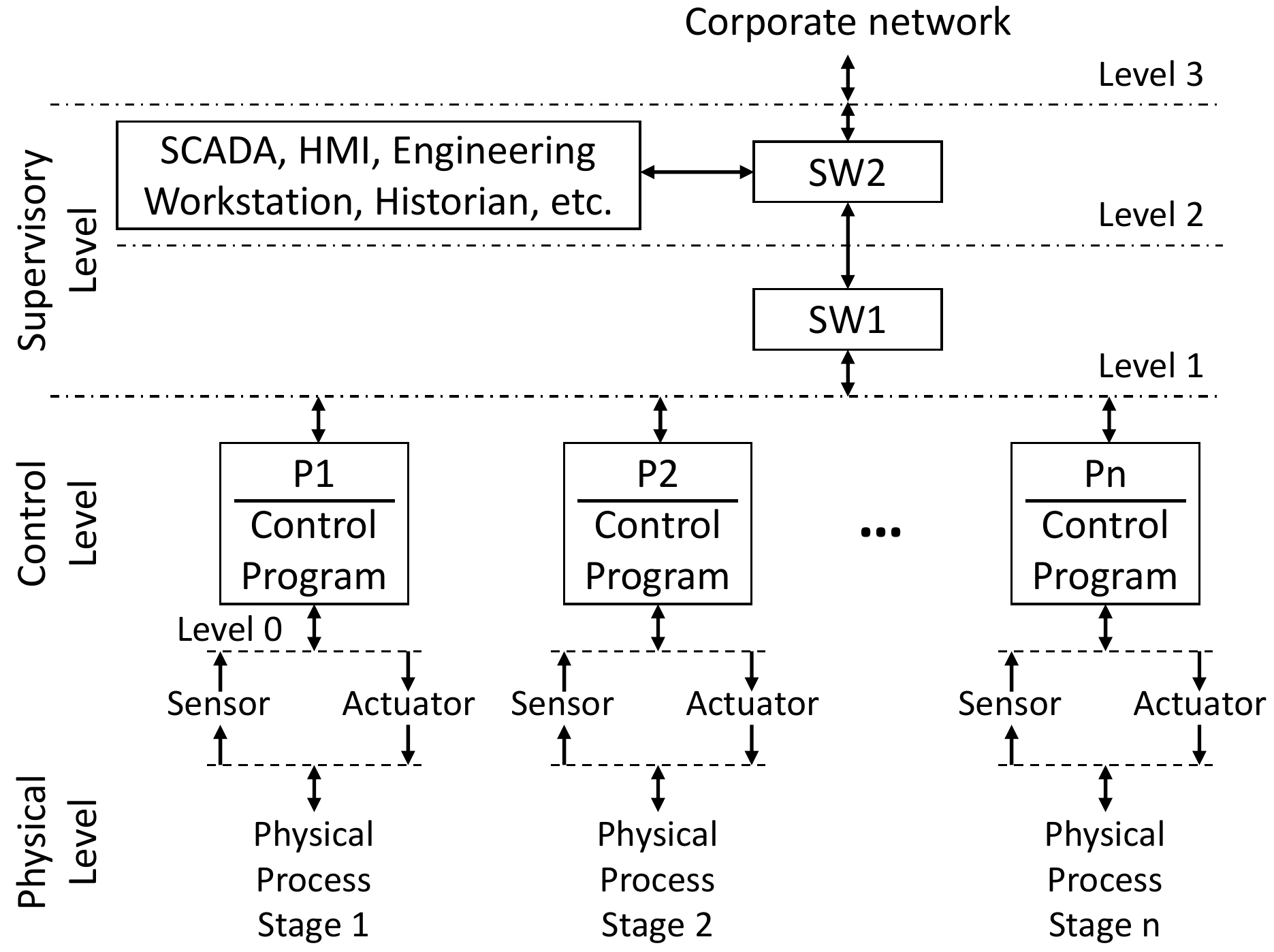}
\caption{Architecture of the control portion of a CPS. $P_1$, $P_2$,...,$P_n$ denote PLCs.}
\label{fig:swatStructure}
\end{figure}

\noindent {Communications}: A multi-layer  network enables communications (as shown in Figure~\ref{fig:swatStructure}) across all components of the network. The ring network at each stage at level~0 enables PLCs to communicate with sensors and actuators at the  corresponding stage. A star network at level~1 enables communications across PLCs, SCADA, HMI and the historian. PLCs communicate with each other through the L1 network, and with centralized Supervisory Control and Data Acquisition (SCADA) system and Human-Machine Interface (HMI), through the Level~2 network. 

\paragraph{Dataset.}
We evaluated \name using data generated by the industrial control network shown in Figure~\ref{fig:swatPhoto}. %
The dataset includes both normal operations to evaluate the false positive rate of \name as well as attacks to evaluate the detection performance of \name.
The attack data, containing data from a total of thirty-six attacks, were generated independently of our work, modeled after Adepu~et~al.~\cite{adepuMathurHASE2016,sridharMathurCOMPSAC2016} and was used in previous works to evaluate ICS security solutions~\cite{gohAdepuTanLee}. 

The dataset contains data collected during seven days of continuous operation of the ICS network. 
It contains $496,800$ data points, each point representing the system state using $53$ features of the system, e.g., sensor values and actuator states.
The sensor data indicates the states of various plant components including tanks, valves, pumps, and meters, as well as data on chemical properties including pH, conductivity, and the Oxidation Reduction Potential (ORP).  




\subsection{\name State Estimation}
We now describe how we evaluated each component of \name's implementation in order to generate the cyber-physical state estimator.

\paragraph{ICS network PLC code consolidation.}
The network consists of six Allen Bradley PLCs using proprietary code and development tools. 
Each PLC was programmed individually using the proprietary Rockwell Automation Studio 5000 development environment. 
In order to perform the code consolidation for each PLC, we first needed to translate the heterogeneously programmed controller projects to a single IEC 61131-3 standard structured text format. 
To do so, we extracted the L5X project files for each PLC~\cite{l5x}. The L5X format is an XML format used for importing/exporting projects to and from the Studio 5000 environment. 
We then built a translation tool, \ltoiec, using an existing \texttt{l5x} Python library\footnote{\url{https://pypi.python.org/pypi/l5x/1.2}} that provides accessors for the XML elements within the L5X files. 
Although the Allen Bradley PLC programming languages conform to the IEC standards, we needed to provide translations for the proprietary 
extensions of each language. 

\paragraph{Physical state estimation.}
\name provides physical state estimation for sensors whose sensor and actuation dependencies are satisfied. 
For instance, we cannot predict the value of a water tank if we do not have access to the corresponding flow rate sensor. 
We provide generic physical state estimators for the water tank level sensors, the flow rate level sensors, as well as the status indicators for pumps and valves.
However, models for other components of the system can be added in future work.

For water tanks, we used the same estimation and threshold values provided in prior analyses of the ICS network platform~\cite{adepu2016using, adepuMathurAsiaCCS2016}. \name implements the following closed-loop state estimation models:
 \begin{center}
$TankLevel = TankLevel + (Inflow-Outflow)*F_c$.
\end{center}
Where \texttt{Inflow} and \texttt{Outflow} are the inflow/outflow rates of the tank and $F_c$ is a conversion constant for the flow rate.

For flow rates, we derived a closed-loop model that incorporates any actuators that may open/close the flow of water:
\begin{center}
$FlowRate = FlowRate * \prod_{n=1}^{N} Actuator_n$
\end{center}
Where $Actuator_n$ represents any pump or value whose value is 0 (for {\em off}) or 1 (for {\em on}). For our analyses, we use the plant invariants that capture the state of the system at any point of time~\cite{adepuMathurAsiaCCS2016}. Each model is then invoked automatically when a particular variable needs to be estimated. In addition to providing generic models for these subsystems, the models for the binary values of the actuator states are automatically generated by our \masterplc executable.

\subsection{Attack Detection}
We were able to successfully detect all attacks enumerated in the attack data set. We further evaluated \name against the record-and-replay attacks enumerated in a previous case study, where sensor values were recorded and replayed back to the HMI to spoof sensor values as was done in the Stuxnet malware~\cite{adepu2016using}. For the non-optimized evaluation, \name had 0 false negatives with a very low false positive rate of 0.36\% for the nominal water tank level deviation threshold in the implementation \emph{without} multi-execution. The false positives were due to the cases mentioned in section \ref{sec:impl},
where an actuator may open/close a tick too early or too late based on our estimated sensor values. 
\begin{figure}[th]
	\centering
	\includegraphics[trim = 24mm 58mm 20mm 56mm, clip, width=1\linewidth]{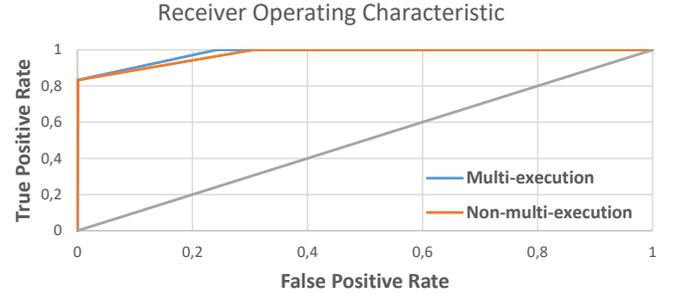}
	\caption{ROC curve for attack detection across varying water tank deviation thresholds for both single and multi-execution analysis.}
	\label{fig:roc}
\end{figure}

\noindent\textbf{False positive pruning.} All false positives were pruned by our error-margin multi-execution implementation for the nominal water tank level deviation threshold. We show the associated ROC curves of varying water tank level deviation thresholds for both the normal execution and the error-margin multi-execution in Figure~\ref{fig:roc}. 
False positives only exist for very small threshold values, i.e., a threshold value that is less than 1mm for the water tank level will obviously result in some false positive rate. The nominal threshold values were based on the threshold values used for state estimation in a previous work~\cite{adepu2016using}. The nominal threshold value of 5mm obtained from our ROC curve confirmed the choice in the previous work. 

\noindent\textbf{Performance.} 
We performed our evaluation on a system equipped with an Intel Core i7-4710MQ Processor at $2.50\,GHz$, $16\,GB$ of RAM running Linux v4.4.0-112-generic.
Running \name on the entire dataset of seven days took $30$ hours for the single threaded deviation-checking, and $51$ hours for the multi-threaded error-margin multi-execution using our current prototype implementation that is not optimized for performance.
This shows that \name can ``keep up'' when running in parallel with the real system even on a desktop-grade computer. 

\paragraph{Memory.} The memory usage of \name was $36\,MB$ on average with a peak memory requirement of $149\,MB$, with multi-execution turned off.
This shows that \name can be used to constantly monitor the system behavior using standard server equipment.
%
%

\paragraph{Communication.} By default, all PLCs communicate with the SCADA system to display the operational process data and to store the operational data in a historian. \name can retrieve its data from the historian causing no communication overhead in the PLC network.

%% file: 8-related-work.tex
\section{Related Work}
\label{sec:related-work}
The previous works on ICS security can be categorized into cyber-physical security mechanisms that are implemented within the ICS controller to enforce code integrity and monitoring solutions that abstract the control of PLCs to verify the overall cyber-physical system. 

\noindent\textbf{Internal CPS control security.}
ECFI~\cite{ecfi} provides a control-flow integrity (CFI) solution for PLCs, where the code running on the PLC is instrumented to validate whether indirect branches follow a legitimate path in the control-flow graph (CFG). 
\name does not need any modifications of the code running on the PLC. In contrast, it monitors the overall behavior of the PLC reflecting its entire software (including the OS).
Furthermore, \name provides context sensitive control-flow checking, i.e., the set of allowed CFG paths is further restricted based on the current system state.

Control-Flow Attestation (C-FLAT) enables a prover device to attest the exact control-flow path of an executed program to a remote verifier~\cite{cflat}. However, 
it cannot be applied to existing systems that do not have the necessary hardware security extensions such as the ARM TrustZone. 
PyCRA\,\cite{ShoukryMYDS15} uses a physical challenge-response authentication to protect active sensing systems against cyber physical attacks.
PyCRA's focus on active sensors, hence it is not applicable to passive sensors nor to actuators, both of which are also common in ICS.
Orpheus~\cite{cheng2017orpheus} monitors the behavior of a program based on executed system calls and checks whether a system call is legitimate in the given context.
The decision is made based on a finite-state machine (FSM) representing the programs system call behavior, i.e., system calls are only allowed to be executed in sequences for which valid transitions exist within the FSM.
Orpheus requires a FSM of the monitored system, which needs to be constructed in a learning phase.
\name does not require such a model of the overall system but only models for the individual subprocesses of the physical system.
Also, Orpheus performs detection on the device and relies on an un-compromised OS and that physical event reports are untampered.
\name does not require any modifications to the monitored devices and does not require a trusted channel to input sensors. 

Zeus~\cite{han2017watch} monitors the control flow of a PLC control program by monitoring the electromagnetic emissions side channels of the PLC by a neural network model.
Such a defense does not protect against data attacks and can further be circumvented via firmware modification attacks. 
Furthermore, Zeus cannot account for verifying other the other networked components as in the \name framework.

State estimation has been used within the PLCs to detect if any of the invariant properties of the system have been violated~\cite{adepu2016using,adepuMathurAsiaCCS2016,adepu2016argus}. This enforcement resides in the application layer of a single PLC, which can be circumvented if the PLC is compromised. Furthermore, the physical invariants and their dependencies are specified manually. \name automatically enforces the checking of discrete-state transitions by analyzing the consolidated PLC code. Similarly, on-device runtime verification has been proposed for PLCs with coupled hypervisors~\cite{garcia2016detecting}. The hypervisor resides above the firmware and relies on the integrity of the PLC control logic.

TSV~\cite{mclaughlin2014trusted} verifies the integrity of any program being loaded onto a PLC by lifting the associated binary to an intermediate language to symbolically execute the program and verify that it is not violating any of the provided infrastructural safety requirements. The safety requirements are enforced within the PLC by extension of the guarantees provided by TSV. Similarly, PLCVerif~\cite{darvas2015plcverif} provides a framework for checking safety properties of PLC code against finite state automata. These solutions are offline analyses that do not provide any runtime guarantees and only verify the control logic application code.

\noindent\textbf{External CPS control security.} Previous works have proposed means of detecting stealthy attacks in the context of ICS.  David et. al.\,\cite{urbinaGiraldoCardenasTippenhauerValenteFaisalRuthsCandellSandberg}  reported on limiting the impact of stealthy attacks on industrial control systems. 
Liu et .al.\,\cite{liuNingReiter,liuNingReiter2011} presented false data injection attacks against state estimation in electric power grids. This work is implemented mainly in a simulation environment, where they are considering stealthy attacks on smart meters. In \name, we also consider stealthy attacks on multiple sensors and actuators on real-time operational data.

Yuqi~et.~al.~\cite{Chen-Poskitt-Sun18a,Chen-Poskitt-Sun16a} proposed an approach for learning physical invariants that combines machine learning with ideas from mutation testing. Initial models are learned using support vector machines. 
These learned models are used for code attestation and identifying standard network attacks. Configuration based intrusion detection system have also been proposed for Advanced Metering Infrastructure~\cite{AliAlshaer}. The AMI behavior is modeled using event logs collected at smart meters. Event logs are modeled using Markov chains and linear temporal logic for the verification of specifications. However, such models depend on the completeness of the training data set used for the learned models. A water control system was modeled using an autoregressive model in order to monitor physics of the system\,\cite{hadziosmanovicSommerZambonHartel}. For distributed systems with complex cyber-physical interdependencies, it is infeasible to assume all discrete states of the system will be traversed. \name automatically contains a discrete-state model of the entire ICS and depends only on the accuracy of the physical state estimation. In a similar vein, the idea of detecting attacks by monitoring physics~\cite{PaulKimballZawodniokRothMcMillinChellappan,ChoudhariRamaprasadPaulKimballZawodniokMcMillinChellappan} of the ICS by using invariants has been applied. However, in these instances the invariants were derived manually based on domain knowledge. \name automatically derives these cyber-physical invariants and significantly reduces the probability of human error during the modelling phase of complex systems.

%% file: 9-discussions.tex
\section{Future Work}
\label{sec:discussion}

In this section we propose possible extensions of \name to improve its security and functionalities.

\paragraph{Simulation interval.}
The current implementation of \name uses a closed loop approach where the system state $s$ after each scan cycle is serving as the basis for the next round.
However, \name can be extended to use state $s$ only after $n$ scan cycles.
This means that the state estimation and multi-execution performed by \name must cover $n$ scan cycles, which could lead to larger errors in the state estimation, which in turn could impact the error-margin multi-execution of \name negatively.
However, this approach can make slowly evolving attacks (see \Cref{sec:security}) even more complicated, further increasing the security of \name.

\paragraph{Automated invariant generation.}
\name cannot only serve as a security solution but can also help improve the functional correctness and safety of an ICS.
Our modeling framework can be used to determine interdependencies of system variables.
This information is useful when programming an ICS as it helps to identify conditions and safety checks that need to be included in the PLCs for the ICS to operate correctly.








%% file: 10-conclusions.tex
\section{Conclusions and Summary}
\label{sec:conclusions}
Industrial control systems (ICS) are ubiquitous and increasingly deployed in critical infrastructures. In fact, recent large-scale cyber attacks (e.g., Stuxnet, BlackEnergy, Duqu to name a few) exploit vulnerabilities in these systems. 
Building a generic defense mechanism against the various ICS attack flavors is highly challenging. However, we observe that all these attacks influence the physics of these devices. As a result, we developed \name, a system that preserves the \emph{Control Behavior Integrity} (CBI) of distributed cyber-physical systems. \name provides real time monitoring for intrusion detection and sensor fault detection by maintaining a cyber-physical state estimation of the system based on a novel control code consolidation generation as well as state estimation equations of the physical processes. \name enforces the correctness of individual controllers in the system by verifying the actuation values being sent from the PLCs as well as the associated changes that propagated through the physical dynamics of the system. We evaluated \name against an enumerated set of attacks on a real water treatment testbed. Our results show that we can detect a wide range of attacks in a timely fashion with zero false positives for nominal threshold values.

%% file: main.bbl
\begin{thebibliography}{10}
\providecommand{\url}[1]{#1}
\csname url@samestyle\endcsname
\providecommand{\newblock}{\relax}
\providecommand{\bibinfo}[2]{#2}
\providecommand{\BIBentrySTDinterwordspacing}{\spaceskip=0pt\relax}
\providecommand{\BIBentryALTinterwordstretchfactor}{4}
\providecommand{\BIBentryALTinterwordspacing}{\spaceskip=\fontdimen2\font plus
\BIBentryALTinterwordstretchfactor\fontdimen3\font minus
  \fontdimen4\font\relax}
\providecommand{\BIBforeignlanguage}[2]{{%
\expandafter\ifx\csname l@#1\endcsname\relax
\typeout{** WARNING: IEEEtranS.bst: No hyphenation pattern has been}%
\typeout{** loaded for the language `#1'. Using the pattern for}%
\typeout{** the default language instead.}%
\else
\language=\csname l@#1\endcsname
\fi
#2}}
\providecommand{\BIBdecl}{\relax}
\BIBdecl

\bibitem{ABE+05}
M.~Abadi, M.~Budiu, U.~Erlingsson, and J.~Ligatti, ``{Control-flow
  Integrity},'' in \emph{Conference on Computer and Communications Security},
  ser. CCS, 2005.

\bibitem{abadi2005control}
------, ``Control-flow integrity,'' in \emph{Proceedings of the 12th ACM
  conference on Computer and communications security}.\hskip 1em plus 0.5em
  minus 0.4em\relax ACM, 2005, pp. 340--353.

\bibitem{ABE+09}
------, ``{Control-flow Integrity Principles, Implementations, and
  Applications},'' \emph{ACM Trans. Inf. Syst. Secur.}, vol.~13, no.~1, Nov.
  2009.

\bibitem{ecfi}
A.~Abbasi, T.~Holz, E.~Zambon, and S.~Etalle, ``{ECFI: Asynchronous Control
  Flow Integrity for Programmable Logic Controllers},'' in \emph{Proceedings of
  the 33rd Annual Computer Security Applications Conference}, ser. ACSAC, 2017.

\bibitem{cflat}
T.~Abera, N.~Asokan, L.~Davi, J.-E. Ekberg, T.~Nyman, A.~Paverd, A.-R. Sadeghi,
  and G.~Tsudik, ``{C-FLAT: Control-Flow Attestation for Embedded Systems
  Software},'' in \emph{Conference on Computer and Communications Security},
  ser. CCS, 2016.

\bibitem{adepuMathurHASE2016}
S.~Adepu and A.~Mathur, ``An investigation into the response of a water
  treatment system to cyber attacks,'' in \emph{Proceedings of the 17th IEEE
  High Assurance Systems Engineering Symposium, Orlando}, January 2016, pp.
  141--148.

\bibitem{adepuMathurAsiaCCS2016}
------, ``Distributed detection of single-stage multipoint cyber attacks in a
  water treatment plant,'' in \emph{Proceedings of the 11th ACM Asia Conference
  on Computer and Communications Security}.\hskip 1em plus 0.5em minus
  0.4em\relax New York, NY: ACM, May 2016, pp. 449--460.

\bibitem{sridharMathurCOMPSAC2016}
------, ``Generalized attacker and attack models for {Cyber-Physical
  Systems},'' in \emph{Proceedings of the 40th Annual International Computers,
  Software \& Applications Conference, Atlanta, USA}.\hskip 1em plus 0.5em
  minus 0.4em\relax Washington, D.C., USA: IEEE, June 2016, pp. 283--292.

\bibitem{adepuMishraMathur}
S.~Adepu, G.~Mishra, and A.~Mathur, ``Access control in water distribution
  networks: A case study,'' in \emph{2017 IEEE International Conference on
  Software Quality, Reliability and Security (QRS)}, July 2017.

\bibitem{adepu2016using}
S.~Adepu and A.~Mathur, ``Using process invariants to detect cyber attacks on a
  water treatment system,'' in \emph{IFIP International Information Security
  and Privacy Conference}, 2016.

\bibitem{adepu2016argus}
S.~Adepu, S.~Shrivastava, and A.~Mathur, ``Argus: An orthogonal defense
  framework to protect public infrastructure against cyber-physical attacks,''
  \emph{IEEE Internet Computing}, vol.~20, no.~5, pp. 38--45, 2016.

\bibitem{AliAlshaer}
\BIBentryALTinterwordspacing
M.~Q. Ali and E.~Al-Shaer, ``Configuration-based ids for advanced metering
  infrastructure,'' in \emph{Proceedings of the 2013 ACM SIGSAC Conference on
  Computer \& Communications Security}, ser. CCS '13.\hskip 1em plus 0.5em
  minus 0.4em\relax New York, NY, USA: ACM, 2013, pp. 451--462. [Online].
  Available: \url{http://doi.acm.org/10.1145/2508859.2516745}
\BIBentrySTDinterwordspacing

\bibitem{antsaklis1993hybrid}
P.~J. Antsaklis, J.~A. Stiver, and M.~Lemmon, ``Hybrid system modeling and
  autonomous control systems,'' in \emph{Hybrid systems}.\hskip 1em plus 0.5em
  minus 0.4em\relax Springer, 1993, pp. 366--392.

\bibitem{l5x}
R.~Automation. (2016) Logix5000 controllers import/exports.
  \url{http://literature.rockwellautomation.com/idc/groups/literature/documents/rm/1756-rm084_-en-p.pdf}.

\bibitem{BBL+13}
Z.~Basnight, J.~Butts, J.~Lopez, and T.~Dube, ``Firmware modification attacks
  on programmable logic controllers,'' \emph{International Journal of Critical
  Infrastructure Protection}, vol.~6, no.~2, pp. 76--84, 2013.

\bibitem{BS15}
M.~Br\"uggemann and R.~Spenneberg, ``Plc-blaster der virus im industrienetz,''
  \url{https://events.ccc.de/congress/2015/Fahrplan/events/7229.html}, 2015.

\bibitem{rop-sparc}
E.~Buchanan, R.~Roemer, H.~Shacham, and S.~Savage, ``{When Good Instructions Go
  Bad: Generalizing Return-oriented Programming to RISC},'' in \emph{ACM
  Conference on Computer and Communications Security}, ser. CCS, 2008.

\bibitem{Chen-Poskitt-Sun16a}
Y.~Chen, C.~M. Poskitt, and J.~Sun, ``{Towards Learning and Verifying
  Invariants of Cyber-Physical Systems by Code Mutation},'' in
  \emph{International Symposium on Formal Methods (FM 2016)}, ser. LNCS, 2016.

\bibitem{Chen-Poskitt-Sun18a}
------, ``Learning from mutants: Using code mutation to learn and monitor
  invariants of a cyber-physical system,'' in \emph{{IEEE} Symposium on
  Security and Privacy (S{\&}P 2018)}.\hskip 1em plus 0.5em minus 0.4em\relax
  {IEEE} Computer Society, 2018, to appear.

\bibitem{cheng2017orpheus}
L.~Cheng, K.~Tian, and D.~D. Yao, ``Orpheus: Enforcing cyber-physical execution
  semantics to defend against data-oriented attacks,'' 2017.

\bibitem{duqu-2011}
E.~Chien, L.~OMurchu, and N.~Falliere, ``{W32.Duqu - The precursor to the next
  Stuxnet},'' Symantic Security Response, Tech. Rep., nov 2011.

\bibitem{ChoudhariRamaprasadPaulKimballZawodniokMcMillinChellappan}
A.~Choudhari, H.~Ramaprasad, T.~Paul, J.~W. Kimball, M.~Zawodniok, B.~McMillin,
  and S.~Chellappan, ``Stability of a cyber-physical smart grid system using
  cooperating invariants,'' in \emph{2013 IEEE 37th Annual Computer Software
  and Applications Conference}, 2013, pp. 760--769.

\bibitem{darvas2015plcverif}
D.~Darvas, E.~Blanco~Vinuela, and B.~Fern{\'a}ndez~Adiego, ``{PLCverif: A tool
  to verify PLC programs based on model checking techniques},'' 2015.

\bibitem{darvasgeneric}
D.~Darvas, I.~Majzik, and E.~B. Vi{\~n}uela, ``Generic representation of plc
  programming languages for formal verification,'' in \emph{Proc. of the 23rd
  PhD Mini-Symposium}.

\bibitem{davis2015cyber}
K.~R. Davis, C.~M. Davis, S.~Zonouz, R.~B. Bobba, R.~Berthier, L.~Garcia, P.~W.
  Sauer \emph{et~al.}, ``A cyber-physical modeling and assessment framework for
  power grid infrastructures,'' 2015.

\bibitem{etigownicyber}
S.~Etigowni, M.~Cintuglu, M.~Kazerooni, S.~Hossain, P.~Sun, K.~Davis,
  O.~Mohammed, and S.~Zonouz, ``{Cyber-Air-Gapped Detection of Controller
  Attacks through Physical Interdependencies}.''

\bibitem{etigowni2016cpac}
S.~Etigowni, D.~J. Tian, G.~Hernandez, S.~Zonouz, and K.~Butler, ``{CPAC:
  securing critical infrastructure with cyber-physical access control},'' in
  \emph{Proceedings of the 32nd Annual Conference on Computer Security
  Applications}.\hskip 1em plus 0.5em minus 0.4em\relax ACM, 2016, pp.
  139--152.

\bibitem{blackenergy}
{F-Secure Labs}, ``{BLACKENERGY and QUEDAGH: The convergence of crimeware and
  APT attacks},'' 2016.

\bibitem{stuxnet-2010}
N.~Falliere, L.~O. Murchu, and E.~Chien, ``{W32.Stuxnet Dossier},'' Symantic
  Security Response, Tech. Rep., Oct. 2010.

\bibitem{rop-avr}
A.~Francillon and C.~Castelluccia, ``{Code Injection Attacks on
  Harvard-architecture Devices},'' in \emph{ACM Conference on Computer and
  Communications Security}, ser. CCS, 2008.

\bibitem{garcia2017hey}
L.~Garcia, F.~Brasser, M.~H. Cintuglu, A.-R. Sadeghi, O.~Mohammed, and S.~A.
  Zonouz, ``{Hey, my malware knows physics! attacking plcs with physical model
  aware rootkit},'' in \emph{Proceedings of the Network \& Distributed System
  Security Symposium, San Diego, CA, USA}, 2017, pp. 26--28.

\bibitem{garcia2014covert}
L.~Garcia, H.~Senyondo, S.~McLaughlin, and S.~Zonouz, ``Covert channel
  communication through physical interdependencies in cyber-physical
  infrastructures,'' in \emph{Smart Grid Communications (SmartGridComm), 2014
  IEEE International Conference on}.\hskip 1em plus 0.5em minus 0.4em\relax
  IEEE, 2014, pp. 952--957.

\bibitem{garcia2016detecting}
L.~Garcia, S.~Zonouz, D.~Wei, and L.~P. de~Aguiar, ``{Detecting plc control
  corruption via on-device runtime verification},'' in \emph{Resilience Week
  (RWS), 2016}.\hskip 1em plus 0.5em minus 0.4em\relax IEEE, 2016, pp. 67--72.

\bibitem{harvey}
L.~A. Garcia, F.~Brasser, M.~Cintuglu, A.-R. Sadeghi, O.~Mohammed, and S.~A.
  Zonouz, ``Hey, my malware knows physics! attacking plcs with physical model
  aware rootkit,'' in \emph{Proceedings of the Network and Distributed System
  Security (NDSS) Symposium}, ser. NDSS, 2017.

\bibitem{gohAdepuTanLee}
J.~Goh, S.~Adepu, M.~Tan, and Z.~S. Lee, ``Anomaly detection in cyber physical
  systems using recurrent neural networks,'' in \emph{2017 IEEE 18th
  International Symposium on High Assurance Systems Engineering (HASE)}, Jan
  2017, pp. 140--145.

\bibitem{hadziosmanovicSommerZambonHartel}
D.~Had\v{z}iosmanovi\'{c}, R.~Sommer, E.~Zambon, and P.~H. Hartel, ``Through
  the eye of the {PLC}: Semantic security monitoring for industrial
  processes,'' in \emph{Proceedings of the 30th Annual Computer Security
  Applications Conference}.\hskip 1em plus 0.5em minus 0.4em\relax ACM, 2014,
  pp. 126--135.

\bibitem{han2017watch}
Y.~Han, S.~Etigowni, H.~Liu, S.~Zonouz, and A.~Petropulu, ``{Watch Me, but
  Don't Touch Me! Contactless Control Flow Monitoring via Electromagnetic
  Emanations},'' in \emph{Proceedings of the 2017 ACM SIGSAC Conference on
  Computer and Communications Security}.\hskip 1em plus 0.5em minus 0.4em\relax
  ACM, 2017, pp. 1095--1108.

\bibitem{dop}
H.~Hu, S.~Shinde, S.~Adrian, Z.~L. Chua, P.~Saxena, and Z.~Liang,
  ``{Data-Oriented Programming: On the Expressiveness of Non-control Data
  Attacks},'' in \emph{IEEE Symposium on Security and Privacy}, ser. S\&P,
  2016.

\bibitem{InoueYCP017}
J.~Inoue, Y.~Yamagata, Y.~Chen, C.~M. Poskitt, and J.~Sun, ``Anomaly detection
  for a water treatment system using unsupervised machine learning,'' in
  \emph{2017 {IEEE} International Conference on Data Mining Workshops, {ICDM}
  Workshops 2017, New Orleans, LA, USA, November 18-21}, 2017, pp. 1058--1065.

\bibitem{swatDataset}
iTrust, ``Swat testbed, dataset and models,''
  \url{https://itrust.sutd.edu.sg/dataset/}, Singapore University of Technology
  and Design.

\bibitem{john2010iec}
K.~H. John and M.~Tiegelkamp, \emph{IEC 61131-3: programming industrial
  automation systems: concepts and programming languages, requirements for
  programming systems, decision-making aids}.\hskip 1em plus 0.5em minus
  0.4em\relax Springer Science \& Business Media, 2010.

\bibitem{junejo2016data}
K.~N. Junejo and D.~K. Yau, ``Data driven physical modelling for intrusion
  detection in cyber physical systems.'' in \emph{Proceedings of the Singapore
  Cyber-Security Conference (SG-CRC)}, 2016, pp. 43--57.

\bibitem{kangAdepuJacksonMathur}
E.~Kang, S.~Adepu, D.~Jackson, and A.~P. Mathur, ``Model-based security
  analysis of a water treatment system,'' in \emph{Proceedings of the 2Nd
  International Workshop on Software Engineering for Smart Cyber-Physical
  Systems}, ser. SEsCPS '16, 2016.

\bibitem{KLM+15}
J.~Klick, S.~Lau, D.~Marzin, J.-O. Malchow, and V.~Roth, ``Internet-facing plcs
  - a new back orifice,'' in \emph{Black Hat USA 2015}, ser. Black Hat USA '15.

\bibitem{KongLiChenSunSunWang}
P.~Kong, Y.~Li, X.~Chen, J.~Sun, M.~Sun, and J.~Wang, ``Towards concolic
  testing for hybrid systems,'' in \emph{FM 2016: Formal Methods}, 2016, pp.
  460--478.

\bibitem{rop-arm}
T.~Kornau, ``{Return Oriented Programming for the ARM Architecture},''
  \url{https://static.googleusercontent.com/media/www.zynamics.com/de//downloads/kornau-tim--diplomarbeit--rop.pdf},
  Tech. Rep., 2009.

\bibitem{LLVM:CGO04}
C.~Lattner and V.~Adve, ``{LLVM: A Compilation Framework for Lifelong Program
  Analysis \& Transformation},'' in \emph{{Proceedings of the 2004
  International Symposium on Code Generation and Optimization (CGO'04)}}, Palo
  Alto, California, Mar 2004.

\bibitem{Liew2017-yh}
D.~Liew, D.~Schemmel, C.~Cadar, A.~F. Donaldson, R.~Z{\"a}hl, and K.~Wehrle,
  ``{Floating-point Symbolic Execution: A Case Study in N-version
  Programming},'' in \emph{Proceedings of the 32Nd {IEEE/ACM} International
  Conference on Automated Software Engineering}, ser. ASE 2017.\hskip 1em plus
  0.5em minus 0.4em\relax Piscataway, NJ, USA: IEEE Press, 2017, pp. 601--612.

\bibitem{lin2018tabor}
Q.~Lin, S.~Adepu, S.~Verwer, and A.~Mathur, ``Tabor: A graphical model-based
  approach for anomaly detection in industrial control systems,'' 2018.

\bibitem{liuNingReiter}
Y.~Liu, P.~Ning, and M.~Reiter, ``False data injection attacks against state
  estimation in electric power grids,'' in \emph{Proceedings of the 16th ACM
  Conference on Computer and Communications Security}, 2009, pp. 21--32.

\bibitem{liu2011false}
Y.~Liu, P.~Ning, and M.~K. Reiter, ``False data injection attacks against state
  estimation in electric power grids,'' \emph{ACM Transactions on Information
  and System Security (TISSEC)}, vol.~14, no.~1, p.~13, 2011.

\bibitem{liuNingReiter2011}
------, ``False data injection attacks against state estimation in electric
  power grids,'' \emph{ACM Transactions on Information and System Security
  (TISSEC)}, vol.~14, no.~1, p.~13, 2011.

\bibitem{mclaughlin2014trusted}
S.~E. McLaughlin, S.~A. Zonouz, D.~J. Pohly, and P.~D. McDaniel, ``A trusted
  safety verifier for process controller code.'' in \emph{Proceedings of the
  Network and Distributed System Security (NDSS) Symposium}, 2014.

\bibitem{moon1994modeling}
I.~Moon, ``Modeling programmable logic controllers for logic verification,''
  \emph{IEEE Control Systems}, vol.~14, no.~2, pp. 53--59, 1994.

\bibitem{SoftBound}
S.~Nagarakatte, J.~Zhao, M.~M. Martin, and S.~Zdancewic, ``{SoftBound: Highly
  Compatible and Complete Spatial Memory Safety for C}, booktitle =
  {Proceedings of the 30th ACM SIGPLAN Conference on Programming Language
  Design and Implementation}, series = {PLDI}, year = {2009},.''

\bibitem{CETS}
------, ``{CETS: Compiler Enforced Temporal Safety for C},'' in
  \emph{Proceedings of the 2010 International Symposium on Memory Management},
  ser. ISMM, 2010.

\bibitem{PalAdepuGoh}
K.~Pal, S.~Adepu, and J.~Goh, ``Effectiveness of association rules mining for
  invariants generation in cyber-physical systems,'' in \emph{18th {IEEE}
  International Symposium on High Assurance Systems Engineering, {HASE} 2017,
  Singapore, January 12-14, 2017}, 2017, pp. 124--127.

\bibitem{PaulKimballZawodniokRothMcMillinChellappan}
T.~Paul, J.~W. Kimball, M.~Zawodniok, T.~P. Roth, B.~McMillin, and
  S.~Chellappan, ``Unified invariants for cyber-physical switched system
  stability,'' \emph{IEEE Transactions on Smart Grid}, vol.~5, no.~1, pp.
  112--120, 2014.

\bibitem{robotlaws}
D.~Quarta, M.~Pogliani, M.~Polino, F.~Maggi, A.~M. Zanchettin, and S.~Zanero,
  ``{An Experimental Security Analysis of an Industrial Robot Controller},'' in
  \emph{IEEE Symposium on Security and Privacy}, ser. S\&P, May 2017.

\bibitem{rop}
R.~Roemer, E.~Buchanan, H.~Shacham, and S.~Savage, ``{Return-Oriented
  Programming: Systems, Languages, and Applications},'' \emph{ACM Trans. Info.
  \& System Security}, vol.~15, no.~1, Mar. 2012.

\bibitem{rrushiquantitative}
J.~Rrushi, H.~Farhangi, C.~Howey, K.~Carmichael, and J.~Dabell, ``A
  quantitative evaluation of the target selection of havex ics malware
  plugin.''

\bibitem{S14}
C.~D. Schuett, ``Programmable logic controller modification attacks for use in
  detection analysis,'' DTIC Document, Tech. Rep., 2014.

\bibitem{ShoukryMYDS15}
Y.~Shoukry, P.~Martin, Y.~Yona, S.~N. Diggavi, and M.~B. Srivastava, ``Pycra:
  Physical challenge-response authentication for active sensors under spoofing
  attacks,'' in \emph{Proceedings of the 22nd {ACM} {SIGSAC} Conference on
  Computer and Communications Security, Denver, CO, USA, October 12-6,
  2015}.\hskip 1em plus 0.5em minus 0.4em\relax {ACM}, 2015, pp. 1004--1015.

\bibitem{stouffer2011guide}
K.~Stouffer, J.~Falco, and K.~Scarfone, ``Guide to industrial control systems
  (ics) security,'' \emph{NIST special publication}, vol. 800, no.~82, pp.
  16--16, 2011.

\bibitem{UmerMathurJunejoAdepu}
M.~A. Umer, A.~Mathur, K.~N. Junejo, and S.~Adepu, ``Integrating design and
  data centric approaches to generate invariants for distributed attack
  detection,'' in \emph{Proceedings of the 2017 Workshop on Cyber-Physical
  Systems Security and PrivaCy, Dallas, TX, USA, November 3}, 2017, pp.
  131--136.

\bibitem{urbinaGiraldoCardenasTippenhauerValenteFaisalRuthsCandellSandberg}
D.~I. Urbina, J.~A. Giraldo, A.~A. Cardenas, N.~O. Tippenhauer, J.~Valente,
  M.~Faisal, J.~Ruths, R.~Candell, and H.~Sandberg, ``Limiting the impact of
  stealthy attacks on industrial control systems,'' in \emph{Proceedings of the
  2016 ACM SIGSAC Conference on Computer and Communications Security}, ser. CCS
  '16.\hskip 1em plus 0.5em minus 0.4em\relax ACM, 2016, pp. 1092--1105.

\bibitem{WangChenSunQin}
J.~Wang, X.~Chen, J.~Sun, and S.~Qin, ``Improving probability estimation
  through active probabilistic model learning,'' in \emph{Formal Methods and
  Software Engineering}, 2017, pp. 379--395.

\bibitem{wang2017should}
J.~Wang, J.~Sun, Q.~Yuan, and J.~Pang, ``Should we learn probabilistic models
  for model checking? a new approach and an empirical study,'' in
  \emph{International Conference on Fundamental Approaches to Software
  Engineering}.\hskip 1em plus 0.5em minus 0.4em\relax Springer, 2017, pp.
  3--21.

\end{thebibliography}
